\documentclass[twocolumn]{aastex62}

\newcommand{\umin}{$U_{\rm min}$}
\newcommand{\kssmc}{S$^3$MC}
\newcommand{\spitzer}{{\it Spitzer}}
\newcommand{\herschel}{{\it Herschel}}
\newcommand{\rv}{$R_{\rm V}$}
\newcommand{\qpah}{$q_{\rm \sc{PAH}}$}
\newcommand{\avqpah}{$\langle q_{\rm \sc{PAH}} \rangle$}
\newcommand{\avqpahref}{$\langle q_{\rm PAH}^{\rm ref} \rangle$}
\newcommand{\fpdr}{$f_{\rm PDR}$}
\newcommand{\ubar}{$\overline{U}$}
\newcommand{\avubar}{$\langle \overline{U} \rangle$}
\newcommand{\halpha}{H$\alpha$}
\newcommand{\hii}{\ion{H}{2}}
\newcommand{\cogas}{$^{12}$CO$(J=1-0)$}

\usepackage{amsmath}
\usepackage{graphicx}
\usepackage{textcomp}
\usepackage{mathtools}
\usepackage{ulem}

\usepackage{hyperref}
\usepackage{cleveref}

\usepackage[encapsulated]{CJK}
\usepackage{ucs}
\usepackage[utf8x]{inputenc}
\shorttitle{PAHs in the MCs}
\shortauthors{Chastenet, Sandstrom et al.}

\begin{document}

\title{The Polycyclic Aromatic Hydrocarbon Mass Fraction\\ on a 10 pc scale in the Magellanic Clouds}


\correspondingauthor{J\'er\'emy Chastenet}
\email{jchastenet@ucsd.edu}

\author[0000-0002-5235-5589]{J\'er\'emy Chastenet}
\affil{Center for Astrophysics and Space Sciences, Department of Physics, University of California, San Diego\\9500 Gilman Drive, La Jolla, CA 92093, USA}

\author[0000-0002-4378-8534]{Karin Sandstrom}
\affiliation{Center for Astrophysics and Space Sciences, Department of Physics, University of California, San Diego\\9500 Gilman Drive, La Jolla, CA 92093, USA}

\author[0000-0003-2551-7148]{I-Da Chiang \begin{CJK*}{UTF8}{bkai}(江宜達)\end{CJK*}}
\affiliation{Center for Astrophysics and Space Sciences, Department of Physics, University of California, San Diego\\9500 Gilman Drive, La Jolla, CA 92093, USA}

\author[0000-0002-2545-1700]{Adam K. Leroy}
\affiliation{Department of Astronomy, The Ohio State University, 4055 McPherson Laboratory, 140 West 18th Ave, Columbus, OH 43210, USA}

\author[0000-0003-4161-2639]{Dyas Utomo}
\affiliation{Department of Astronomy, The Ohio State University, 4055 McPherson Laboratory, 140 West 18th Ave, Columbus, OH 43210, USA}

\author[0000-0001-6118-2985]{Caroline Bot}
\affiliation{Observatoire astronomique de Strasbourg, Université de Strasbourg, CNRS, UMR 7550, 11 rue de l\textquotesingle Universit\'e, F-67000 Strasbourg, France}

\author[0000-0001-5340-6774]{Karl D. Gordon}
\affiliation{Space Telescope Science Institute, 3700 San Martin Drive, Baltimore, MD, 21218, USA}
\affiliation{Sterrenkundig Observatorium, Universiteit Gent, Gent, Belgium}

\author[0000-0002-0846-936X]{Bruce T. Draine}
\affiliation{Princeton University Observatory, Peyton Hall, Princeton, NJ 08544-1001, USA}

\author{Yasuo Fukui}
\affiliation{Institute for Advanced Research, Nagoya University, Furo-cho, Chikusa-ku, Nagoya 464-8601, Japan}
\affiliation{Department of Physics, Nagoya University, Furo-cho, Chikusa-ku, Nagoya 464-8601, Japan}

\author{Toshikazu Onishi}
\affiliation{Department of Physical Science, Graduate School of Science, Osaka Prefecture University, 1-1 Gakuen-cho, Naka-ku, Sakai, Osaka 599-8531, Japan}

\author[0000-0002-2794-4840]{Kisetsu Tsuge}
\affiliation{Department of Physics, Nagoya University, Furo-cho, Chikusa-ku, Nagoya 464-8601, Japan}


\begin{abstract}
\noindent We present maps of the dust properties in the Small and Large Magellanic Clouds (SMC, LMC) from fitting Spitzer and Herschel observations with the \citet{DL07} dust model. We derive the abundance of the small carbonaceous grain (or polycyclic aromatic hydrocarbon; PAH) component. The global PAH fraction (\qpah, the fraction of the dust mass in the form of PAHs) is smaller in the SMC (1.0$^{+0.3}_{-0.3}$~\%) than in the LMC (3.3$^{+1.4}_{-1.3}$~\%).
We measure the PAH fraction in different gas phases (\hii\ regions, ionized gas outside of \hii\ regions, molecular gas, and diffuse neutral gas). 
\hii\ regions appear as distinctive holes in the spatial distribution of the PAH fraction. In both galaxies, the PAH fraction in the diffuse neutral medium is higher than in the ionized gas, but similar to the molecular gas.
Even at equal radiation field intensity, the PAH fraction is lower in the ionized gas than in the diffuse neutral gas. 
We investigate the PAH life-cycle as a function of metallicity between the two galaxies.
The PAH fraction in the diffuse neutral medium of the LMC is similar to that of the Milky Way ($\sim4.6~\%$), while it is significantly lower in the SMC. 
Plausible explanations for the higher PAH fraction in the diffuse neutral medium of the LMC compared to the SMC include: a more effective PAH production by fragmentation of large grains at higher metallicity, and/or the growth of PAHs in molecular gas.
\end{abstract}

\keywords{ISM: abundances -- (ISM:) dust -- (galaxies:) Magellanic Clouds}
%globular clusters: individual(\objectname{NGC 6397},
%\object{NGC 6624}, \objectname[M 15]{NGC 7078},
%\object[Cl 1938-341]{Terzan 8})

\section{Introduction}
Dust grains have a major impact on the energy balance and chemistry of the interstellar medium (ISM), and therefore are critical to the evolution of a galaxy.
Acting as a favored surface for H$_2$ formation, they are a key agent in the chemical balance of the ISM \citep[][]{LePage09,LeBourlot12,Bron14}. 
Dust grains are also an efficient heat source for the ISM through photoelectric heating, which is the main mechanism that heats the neutral gas \citep{Wolfire95}.
The effectiveness with which dust grains play these key roles in the ISM depends on their intrinsic properties, e.g. size, charge, and chemical composition. 
To understand the effects of dust grains on the ISM, we need to understand their properties.

In this regard, the smallest grains are of particular interest. The small carbonaceous grain component is thought to be in the form of polycylic aromatic hydrocarbons \citep[PAHs;][]{Leger84, Allamandola85, Allamandola89}. 
PAHs play an important role in the photo-electric heating of the ISM, the efficiency of which depends on their UV absorption cross-section and grain charge \citep{Bakes94, Weingartner01pe}. 

PAHs are also widely considered to be responsible for the mid-IR (MIR) emission features.
Their emission dominates the MIR through broad emission bands at 3.3, 6.2, 7.7, 11.3, and 17~$\mu$m.
% $\mu$m.
These prominent features trace the vibrational modes of the C--C, C--C--C, and C--H bonds in PAHs, and can be used to probe the ionization and size distribution of the PAH population \citep[for a review, see][]{Tielens08}.
Because the mid-IR bands from PAHs are preferentially excited by higher energy photons,
they are often considered as a tracer of star formation \citep{Peeters04}. 
The intensity of some of the emission bands even allows for detection at high redshift, and can be used to determine the star formation earlier in the history of the Universe \citep{Sajina09, Siana09, Shipley16}. At intermediate redshifts, they are found to trace molecular gas \citep[][]{Cortzen19}.
PAHs are also candidate carriers of the ``2175~\AA\ bump'', seen in extinction \citep{Mathis94, Steglich10}. This intriguing feature shows variations both in width and intensity between lines of sight, between galaxies and within the same galaxy \citep{Gordon03}. To understand ISM thermal balance, mid-IR emission and UV absorption in galaxies, we must understand the life cycle of PAHs.

Studies have found evidence of changes in PAH properties as a function of galaxy properties, particularly a PAH deficiency at low metallicity  \citep[][]{Engelbracht05, Madden06, Draine07, Galliano08, Sandstrom10, Paradis11, RemyRuyer15}. The intensity of the MIR features decreases in these environments relative to the total IR emission. This suggests that there is a change in the dust composition, with a lower abundance of the grains responsible for the MIR emission, compared to the larger grains, emitting primarily at far-infrared (FIR) wavelengths. In the \citet{DL07} dust model, the PAH fraction is defined as the fraction of the total dust mass in grains with less than 10$^3$ carbon atoms, and is hereafter labeled \qpah. 
The Galactic diffuse neutral medium \qpah\ lies around 4.6~\%\footnote{It is difficult to estimate a systematic uncertainty on \qpah. It depends on the physics of PAHs (e.g., UV, optical and IR cross sections, broad continuum PAH emission, which show a large scatter in their theoretical values), and modeling assumption (e.g., starlight spectral shape).} \citep{LD01, WD01}.

\citet{Draine07} measured \qpah\ in the SINGS galaxy sample and found a wide range of \qpah, from $\sim 0.5~\%$ in dwarf galaxies, up to almost $\sim 5~\%$ in spiral galaxies.
Their results suggested a dependence between the PAH fraction and the metallicity of a galaxy, where \qpah\ drops at lower metallicity. Some studies have suggested a threshold in metallicity, around $12+{\rm log(O/H) \sim 8.0-8.2}$, at which the PAH abundance varies drastically \citep{Draine07}. On the other hand, \citet[][]{RemyRuyer15} find a power-law relationship between metallicity and \qpah, rather than a step-function.
There are several hypotheses for the dearth of PAHs in low-metallicity environments. PAHs could be exposed to more intense and/or harder far-UV radiation fields due to the overall decrease in dust shielding, and suffer from a more efficient selective photo-destruction \citep[e.g.][]{Madden06}. PAHs could also form in the dense ISM \citep[][]{Zhukovska16}, and this process could be less effective at lower metallicities. Or the low PAH abundance could be the sign of a lower efficiency of forming PAH-like dust, due to particular stellar evolution at low metallicity \citep[][]{Galliano08}.

In order to study both the metallicity trend, and disentangle the various local influences on the PAH fraction in a galaxy, we focus on two of the closest galaxies, the Magellanic Clouds (MCs). The Small Magellanic Cloud (SMC) lies at about 62~kpc \citep[][]{Graczyk14}, and has a metallicity of $\sim 1/5~Z_\odot$ \citep{RusselDopita92}. The Large Magellanic Cloud is closer, at about 50~kpc \citep{Walker12}, and has a higher metallicity, of $\sim 1/2~Z_\odot$ \citep{RusselDopita92}. 
The MCs therefore represent attractive targets for detailed studies of dust properties \citep[e.g.][]{Leroy07, Bernard08, Paradis09, Bot10, Israel10, Sandstrom10, Galliano11, Paradis11, Sandstrom12, Gordon14, RomanDuval14, Tchernyshyov15}.
Their respective metallicities bracket a threshold at which PAH properties are thought to vary significantly \citep{Draine07}. Here, we present a study of the PAH fraction across both MCs at 10 pc resolution, with the same dust grain model, in order to compare to the MW and other nearby galaxies. Thanks to the MCs proximity, we can resolve ISM structures, such as \ion{H}{2} regions, in the FIR. This allows us to provide detailed maps of 10~pc scale PAH abundance, and investigate its variation as a function of the dominant ISM phase.
Throughout this paper, we assume a constant metallicity across each galaxy. However, recent studies by \citet[][]{Fukui2017} and \citet[][]{Tsuge19} suggested that the \ion{H}{1} ridge of the LMC, south of the star forming complex 30~Doradus, is mainly SMC-stripped gas from an colliding event.

Our paper is laid out as follows: we first describe the data used in this study, in Section \ref{SecData}. The dust grain model used to fit the dust IR emission is detailed in Section \ref{SecMethodo}, before giving the results of the fit, and showing the variations of the PAH fraction with environment, in Section \ref{SecResults}. Finally, Section \ref{SecDiscussion} is dedicated to discussion and interpretation of the PAH abundance variations in the MCs.

\section{Data}
\label{SecData}
\subsection{Infrared}
We combine observations from the {\it Spitzer} Space Telescope \citep{Werner04} and the {\it Herschel} Space Observatory \citep{Pilbratt10} to cover the mid- through far-IR spectral energy distribution (SED).
In the mid-infrared (MIR), we use observations from the {\it Spitzer} Legacy program SAGE (Surveying the Agents of the Galaxy), which observed both the LMC \citep[SAGE-LMC;][]{Meixner06} and the SMC \citep[SAGE-SMC;][]{Gordon11}. The final images produced by SAGE-SMC include deeper measurements in the main star forming regions of the SMC, from the \kssmc\ program \citep{Bolatto07}. 
In the far-infrared (FIR) and sub-millimeter range, the {\it Herschel} Key Project HERITAGE \citep[the Herschel Inventory of the Agents of Galaxy Evolution;][]{Meixner13,Meixner13err} surveyed both clouds. This leads to a total of 11 photometric bands included in our study, at 3.6, 4.5, 5.8, 8.0 \citep[\spitzer, IRAC;][]{Fazio04}, 24, 70 \citep[\spitzer, MIPS;][]{Rieke04}, 100, 160\footnote{We use the PACS160 instead of MIPS160 band, for the sake of resolution.} \citep[\herschel, PACS;][]{Poglitsch10}, 250, 350, and 500~$\mu$m \citep[\herschel, SPIRE;][]{Griffin10}.

The goal of this study is to model dust emission from the MIR to the sub-millimeter range. However, there is a significant contribution from stars in the \spitzer\ IRAC bands. 
For the average stellar populations in a pixel, we can account for stellar emission with a simple assumption of a 5,000~K blackbody \citep[][and see next Section]{Draine07}, but due to the proximity of the Magellanic Clouds, occasionally a resolution element is dominated by a very bright source that is not well modeled by a blackbody. In particular, this can be young stellar objects or evolved stars \citep{Woods11,Jones17}. To avoid contamination by these sources, we mask out the bright point sources that show up in the short wavelength bands.
We perform our own masking of the brightest sources in these bands, in two steps. A first set of sources is simply chosen by looking at the images, and selecting the brightest stars. Second, to remove as much of the contaminating point sources as possible, we perform a fit with the un-masked images. Using these fit results, we mask stars where there is an evident bias in the parameter maps due to a point source in the image. We mask all the selected sources by replacing the value within a small radius from the source with the averaged value of the local diffuse emission.

Prior to fitting we perform several additional steps of image processing.
We correct the IRAC and SPIRE images with extended source factors, as suggested by the IRAC Instrument Handbook\footnote{\url{https://irsa.ipac.caltech.edu/data/SPITZER/docs/irac/iracinstrumenthandbook/29/}} and the KINGFISH User Guide\footnote{\url{http://irsa.ipac.caltech.edu/data/Herschel/KINGFISH/docs/KINGFISH_DR3.pdf}}.
Then, all observations are convolved to the SPIRE~500 resolution ($\sim 36$'') using the \citet{Aniano11} convolution kernels. 
Even though the clouds are at relatively high Galactic latitude, the observations still suffer from a non-negligible emission from the Milky-Way cirrus. We follow \citet[][]{Gordon14} and \citet{Chastenet17} to remove this foreground emission. We convert the foreground \ion{H}{1} MW column density map to a dust column density using coefficients $1.47\times10^{-3}$, $9.32\times10^{-4}$, $1.34\times10^{-2}$, $4.28\times10^{-2}$, $2.53\times10^{-2}$, $2.63\times10^{-1}$, 1.36, 1.07, 1.85, 1.20, 0.62, 0.25~MJy/sr ($10^{20}$~\ion{H}{1}/cm$^{2}$)$^{-1}$ from 3.6 to 500~$\mu$m respectively, and subtract the foreground cirrus.
We then perform a background subtraction to get rid of residual emission from the cosmic infrared background, zodiacal light and mosaicing offsets. 
Background regions are selected by eye: we visually identify portions of the images where we can avoid  contamination from dust emission from the target galaxies. In the LMC, these are chosen to be at the edges of the images. In the SMC, we avoid the SMC Bar and wing to select the background pixels. We fit and subtract a tilted plane, which removes the gradient across the background.
All images were then projected with the final pixel grid sampling the point spread function with approximately independent pixels, that is l$_{\rm pixel} \sim 42$'', which corresponds to a pixel size of $\sim 12$~pc in the SMC and $\sim 10$~pc in the LMC.
After the final projection, a background covariance matrix $\mathcal{C}_{\rm bkg}$ was constructed from the background pixels, to quantify the correlations between noise in different photometric bands \citep{Gordon14}.

\subsection{Additional data}
In this study, we are interested in possible correlations between the fitted dust properties and other components of the ISM. 
We use the SHASSA survey \citep[the Southern H-Alpha Sky Survey Atlas;][]{Gaustad01} to investigate the spatial distribution of the ionized gas from \halpha~emission\footnote{The maps are in units of dR.\newline1~R = 10$^6$/4$\pi$~photons cm$^{-2}$ s$^{-1}$ sr$^{-1}$.\newline1~R = $5.661 \times 10^{-18}$ erg s$^{-1}$ cm$^{-2}$ arcsec$^{-2}$}. We use the smoothed maps of the LMC (field 013), and SMC (field 510), at 4\arcmin\ angular resolution. The maps were projected onto the final pixel grid of our data set, thus oversampling the point spread function of the \halpha\ emission data.

We use $^{12}$CO$(J=1-0)$ maps from the NANTEN \citep{Fukui91} survey of the SMC \citep{Mizuno01smc} and the LMC \citep{Fukui08lmc} to trace the spatial distribution of the molecular gas ($\sim 3'$ resolution). In Section \ref{SecPAHsgas}, we will use a 3-$\sigma$ detection threshold in CO integrated intensity to define the ``molecular gas phase''. We determine this threshold by computing the standard deviation $\sigma$ in a region where we do not find any detections by eye. We will consider the molecular gas phase as every pixel above 3-$\sigma$. In the SMC, we find a 3-$\sigma$ value of 0.3 K~km~s$^{-1}$, and 0.75~K~km~s$^{-1}$ in the LMC.
Assuming $\alpha_{\rm CO}^{\rm SMC}=76~{\rm M_\odot\ pc^{-2}\ (K\ km\ s^{-1})^{-1}}$, and $\alpha_{\rm CO}^{\rm LMC}=10~{\rm M_\odot\ pc^{-2}\ (K\ km\ s^{-1})^{-1}}$ \citep{Jameson16}, this corresponds to a molecular gas surface density of 22.8~${\rm M_\odot\ pc^{-2}}$ in the SMC, and 7.5$~{\rm M_\odot\ pc^{-2}}$ in the LMC.

\section{Spectral Energy Distribution Fitting Methodology}
\label{SecMethodo}
We use the \citet{DL07} model to fit the dust emission in our data set, from 3.6 to 500~$\mu$m \citep[with updates similar as those from][]{Aniano12}.
The model has a size distribution of grains divided into a carbonaceous and a silicate component. The fraction of the dust grain mass made up of PAHs with less than 10$^3$ carbon atoms is given by the \qpah\ parameter. Like \citet[][]{Draine07} and \citet[][]{Aniano12} we fit the SED using the MW \rv~$=3.1$ grain model. This model has been found to provide good fits in low metallicity conditions by \citet[][]{Draine07} and \citet[][]{Sandstrom10} and allows us to include \qpah\ as a fit parameter, which is critical to our study.

In each pixel $j$, the dust is heated by a range of radiation field intensities described by the parameter $U$, a dimensionless factor scaling the \citet{Mathis83} 10~kpc Milky Way interstellar radiation field. In each pixel, a fraction $(1-\gamma)$ of the dust grain mass is heated by a radiation field of intensity \umin. The remaining fraction $\gamma$ is heated by a power-law distribution of $U_{\rm min} \leq U \leq {\rm U_{max}}$  \citep[see equations 8, 9 and 10 of][]{Aniano12}, with a power-law exponent $\alpha$. In this study, we fix ${\rm U_{max}}=10^7$, and $\alpha=2$. 
In the end, we have 5 free parameters: the minimum radiation field \umin, $\gamma$ the fraction of the dust mass heated by the power-law distribution of radiation fields, the PAH fraction \qpah, the dust surface density $\Sigma_{\rm d}$, and the scaling parameter of stellar surface brightness, $\Omega_*$, which adjusts a 5,000~K blackbody to match the observed starlight continuum in the shortest wavelength bands. See Table~\ref{TabParams} for the boundaries of each parameter (Range), and the sampling (Step). The parameter values ensure a sampling fine enough to resolve the 1-D likelihood functions and were determined after several iterations of the fit.

\renewcommand{\arraystretch}{1.2}
\begin{deluxetable}{llll}
\caption{Fitting parameters}
    \centering
    \tablehead{
    \colhead{Parameter} & \colhead{Unit} & \colhead{Range} & \colhead{Step}} 
    \startdata
    $U_{\rm min}$ &  & [0.1, 50] & Uneven spacing\tablenotemark{a}\\
    log(q$_{\rm PAH}$) & \% & [-1.0, 0.88]\tablenotemark{b} & 0.0725\\
    log$_{10}$($\gamma$) & - & [-3.3, 0] & 0.1 \\
    log$_{10}$($\Sigma_{\rm d}$) & M$_\odot$/pc$^2$ & [-2.0, 0.7] & 0.15\\
    log$_{10}$($\Omega_*$) & - & [-2.0, 2.7] & 0.15\\
    \enddata
\tablenotetext{a}{$U_{\rm min} \in $ \{0.1, 0.12, 0.15, 0.17, 0.2, 0.25, 0.3, 0.35, 0.4, 0.5, 0.6, 0.7, 0.8, 1.0, 1.2, 1.5, 1.7, 2.0, 2.5, 3.0, 3.5, 4.0, 5.0, 6.0, 7.0, 8.0, 10.0, 12.0, 15.0, 17.0, 20.0, 25.0, 30.0, 35.0, 40.0, 50.0\}.}
\tablenotetext{b}{We also include q$_{\rm PAH}=0$.}
\label{TabParams}
\end{deluxetable}

Using the fitted parameters, we calculate, in each pixel $j$, \ubar\ and \fpdr, as described by \citet{Aniano12}. \ubar\ measures the dust-mass-weighted average radiation field intensity, generally given by:
\begin{equation}
    \begin{split}
    &\overline{U_j} = (1-\gamma_j)\ U_{{\rm min,}j} + \gamma_j \\
    &\times
        \begin{dcases}
        \left (\frac{\alpha_j-1}{\alpha_j-2} \right )\ \left ( \frac{{\rm U^{2-\alpha_j}_{max}} - U_{{\rm min,}j}^{2-\alpha_j}}{{\rm U^{1-\alpha_j}_{max}} - U_{{\rm min,}j}^{1-\alpha_j}} \right ),\; {\rm if\ \alpha \neq 1,\ \alpha \neq 2};\\
        U_{{\rm min,}j} \ \frac{{\rm ln(U_{max}}/U_{{\rm min}j})}{1 - U_{{\rm min,}j}/{\rm U_{max}} },\; {\rm if\ \alpha = 2}.
        \end{dcases}
    \end{split}
\end{equation}
\fpdr~is the fraction of the dust luminosity produced by regions where $U \geq {\rm U_{PDR}} = 10^2$:
\begin{equation}
    f_{\rm PDR} = L_{\rm PDR} / L_{\rm dust}
\end{equation}
with
\begin{equation}
\begin{split}
    L_{{\rm PDR,}j} =& P_{0,j}(q_{\rm PAH}) \, M_{{\rm d,}j} \, \gamma_j \,
        \frac{{\rm ln}({\rm U_{max}/U_{PDR}})}{U^{-1}_{{\rm min,}j} - {\rm U^{-1}_{max}}}, \\
    L_{{\rm dust,}j} =& P_{0,j}(q_{\rm PAH}) \, M_{{\rm d,}j} \, \overline{U_j}. 
\end{split}
\end{equation}
Here $P_0(q_{\rm PAH})$ is the power radiated per unit dust mass, when $U=1$, M$_{\rm d}$ is the dust mass. 

Aside from their pixel-to-pixel distribution, we measure galaxy-average values of the fitted parameters, weighted according to the dust mass or luminosity distribution. Following \citet{Aniano12}, we define the dust-mass averaged value of the mass fraction of PAHs,
\begin{equation}
\langle q_{\rm PAH} \rangle = \frac{\sum^N_j \ q_{{\rm PAH,}j} \, M_{{\rm d,}j}}{\sum^N_j \  M_{{\rm d,}j}},
\label{EquAvqpah}
\end{equation}
and similarly, the dust-mass averaged starlight intensity,
\begin{equation}
\langle \overline{U} \rangle = \frac{\sum^N_j \ \overline{U}_{j} \, M_{{\rm d,}j}}{\sum^N_j \  M_{{\rm d,}j}}.
\label{EquAvUbar}
\end{equation}
The average value of \fpdr, i.e. dust luminosity weighted, is expressed as
\begin{equation}
\langle f_{\rm PDR} \rangle = \frac{\sum^N_j \ {L_{\rm PDR,}j} }{\sum^N_j \  L_{{\rm dust,}j}}.
\end{equation}
We assume that $P_0(q_{\rm PAH})$ variations are small enough pixel-to-pixel to be negligible \citep{Aniano12}.

\subsection{Fitting and uncertainties}
\begin{figure}
    \centering
    \includegraphics[width=0.49\textwidth]{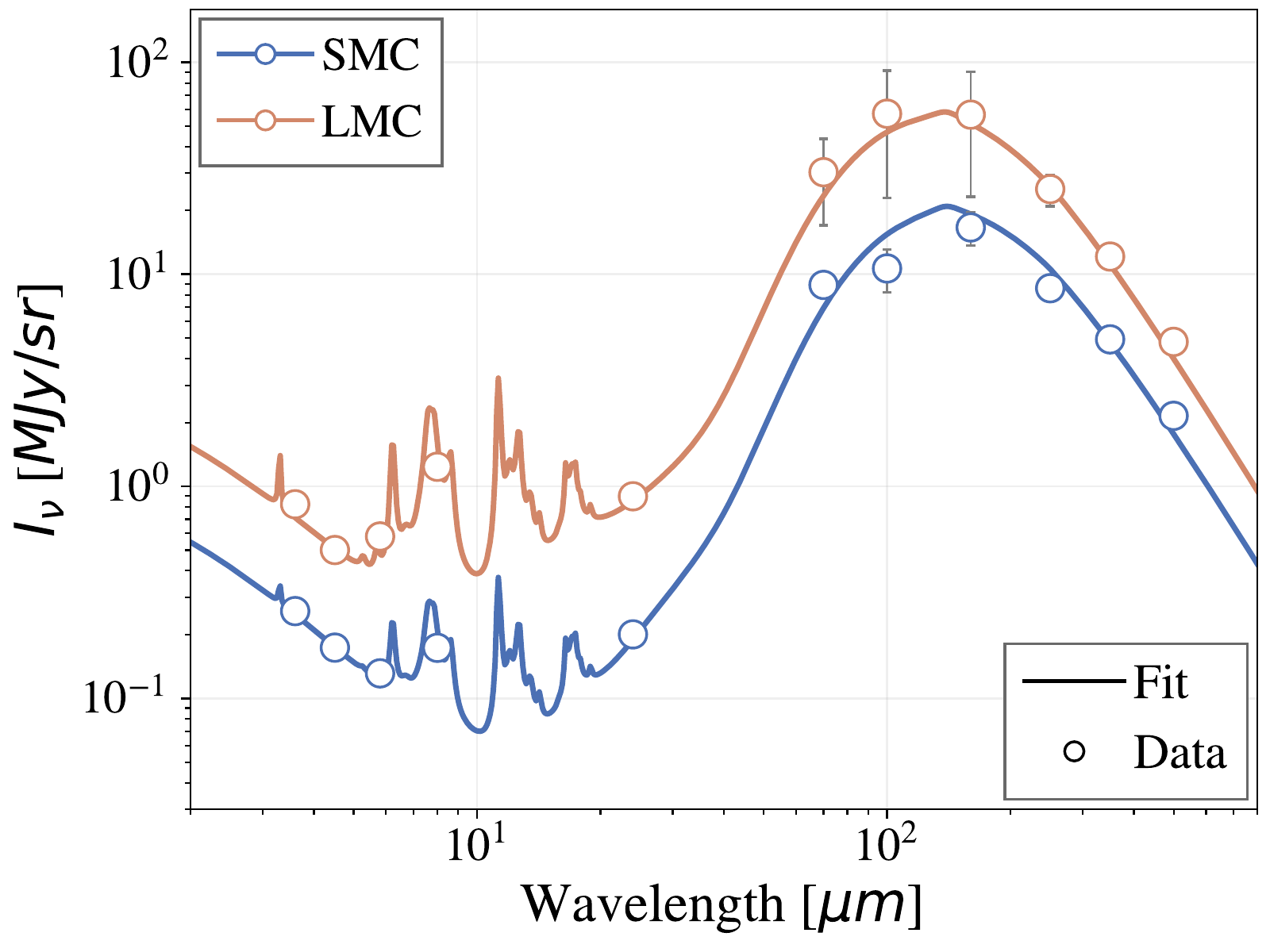}
    \caption{\small{Examples of the fitting results in two pixels in the diffuse ISM of the SMC (blue) and the LMC (orange). The circles mark the data SEDs with their 3-$\sigma$ errors, and the solid lines show the best fit model, in each case.}}
    \label{FigFits}
\end{figure}

In Figure \ref{FigFits} we show two examples of the data in the SMC and the LMC fit by the best model (i.e. maximum likelihood), in two pixels of the diffuse ISM. The error bars at short wavelengths show that the errors are small enough to strongly constrain \qpah.
Residuals show that the fits are good in both galaxies at short wavelengths, especially 8~$\mu$m. At long wavelengths, the fractional residuals are mostly negative, indicating that the model generally overestimates the data (residuals peak at less than 10~\%).

The fitting is done with the DustBFF tool \citep[][]{Gordon14}, which determines the n-dimensional {\it a posteriori} likelihood distribution of the parameters. It is a Bayesian fitting tool that uses flat priors on the parameter distributions, and a covariance matrix, built from background pixels and instrument errors, to propagate uncertainties.
We use the same calibration errors on \spitzer~and \herschel~instruments as those in \citet{Gordon14} and \citet[][]{Chastenet17}.
The diagonal elements of the covariance matrix include both uncorrelated (or statistical) and correlated errors, and the non-diagonal elements measure only correlated errors between the photometric bands.  We refer the reader to \citet{Gordon14} for further details on the covariance matrix and the DustBFF fitting.

We use the 5-dimensional (for 5 fitting parameters) likelihood function to build realizations of the parameter maps: a realization is a sample of the 5-dimensional grid, weighted by the likelihood function. The realizations render noise properties more accurately than finding only the maximum of the likelihood function. To measure values like \avqpah\ and their associated 16$^{\rm th}$ and 84$^{\rm th}$ percentiles, we prefer the realization method over the expectation value \citep[][]{Gordon14,Chiang18,Utomo19}, since it draws samples without marginalizing the likelihood over all but one parameter dimension. 
Since we compute weighted-values with both \qpah\ and $\Sigma_{\rm d}$, it is better to use the full-dimension likelihood distribution to avoid losing information. We build a large number of realization maps for each parameter simultaneously (both the fitted parameters and the calculated parameters like \ubar\ and \fpdr), and use them to calculate the mean in each pixel. This is what is presented in Figures \ref{FigResultsSMC} and \ref{FigResultsLMC}.

From  the realizations of a single pixel or group of pixels, we can determine its statistical error by measuring the standard deviation from a large number of realizations. When combining many pixels together we can propagate these uncertainties to calculate the error on the mean values.
To represent the intrinsic scatter of a parameter within a specific region of the galaxy (e.g. Figures \ref{FigMCsGasCorr} or \ref{FigUbarqPAH}), we use the 16$^{\rm th}$ and 84$^{\rm th}$ percentiles of the dust mass (or dust luminosity for \fpdr) weighted distribution of said parameter. For all averaged parameters, e.g. \avqpah, quoted in the rest of the paper, the statistical uncertainties are generally very small, due to the large numbers of pixels begin averaged together.
We instead list the $\pm$ the 16$^{\rm th}$ and 84$^{\rm th}$ percentiles of the dust mass- or luminosity-weighted distributions. 

\section{Results}
\label{SecResults}
\subsection{Fitting parameter maps}
\begin{figure*}
    \centering
    \includegraphics[trim={0 7cm 0cm 1cm},clip,width=0.99\textwidth]{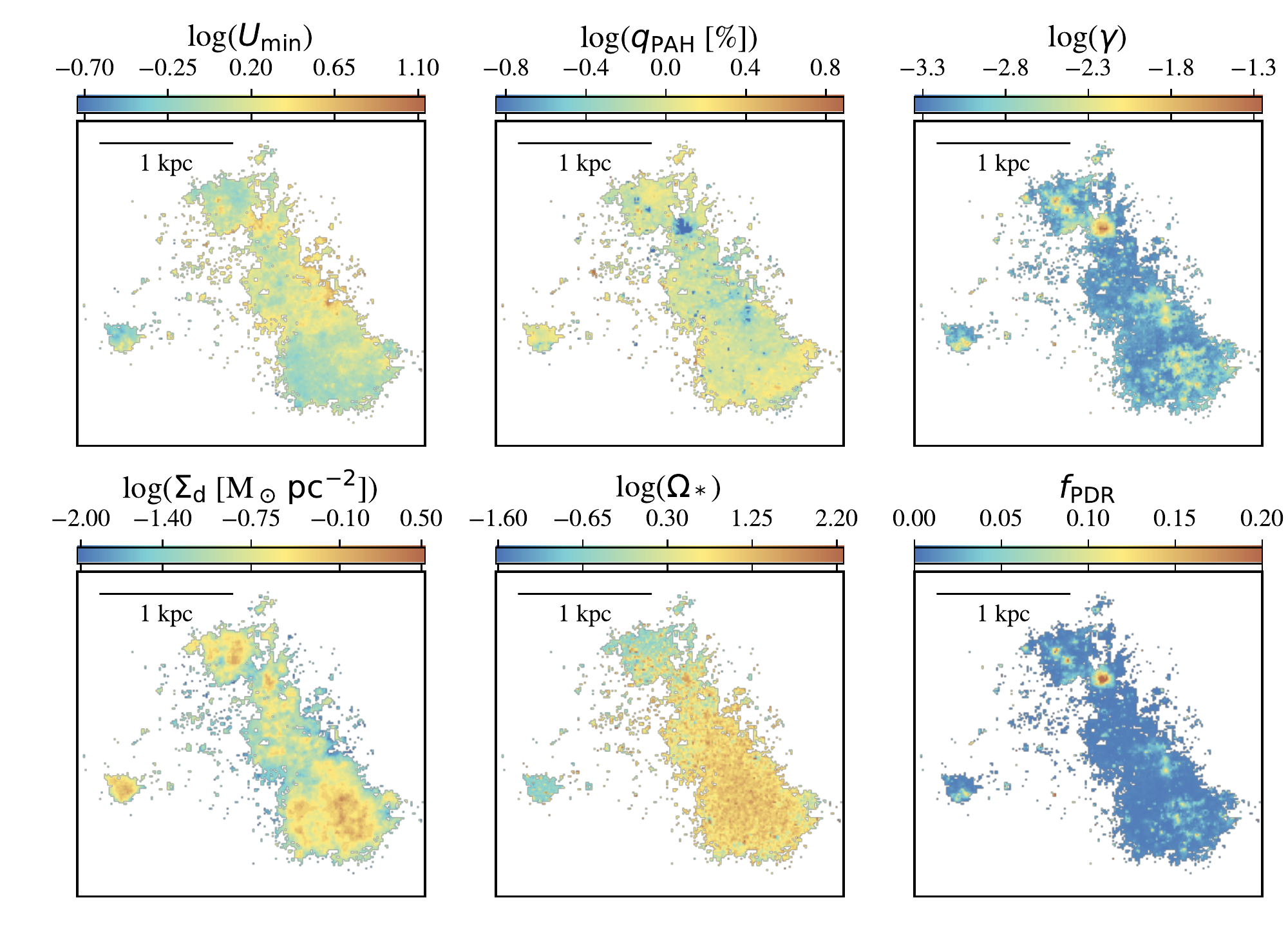}
    \caption{\small{Results of the fitting in the SMC for the minimum radiation field, \umin, the PAH mass fraction, \qpah\ (see Figure~\ref{FigMCsqPAHImg}, and Section~\ref{SecPAHinMC}), the weight of dust mass heated by a power-law combination of radiation field, $\gamma$, the total dust surface density, $\Sigma_{\rm d}$, the scaling of stellar surface brightness, $\Omega_*$, and the derived fraction of dust luminosity where U $> 10^2$, \fpdr.}}
    \label{FigResultsSMC}
\end{figure*}

\begin{figure*}
    \centering
    \includegraphics[trim={0 7cm 0 0},clip,width=0.99\textwidth]{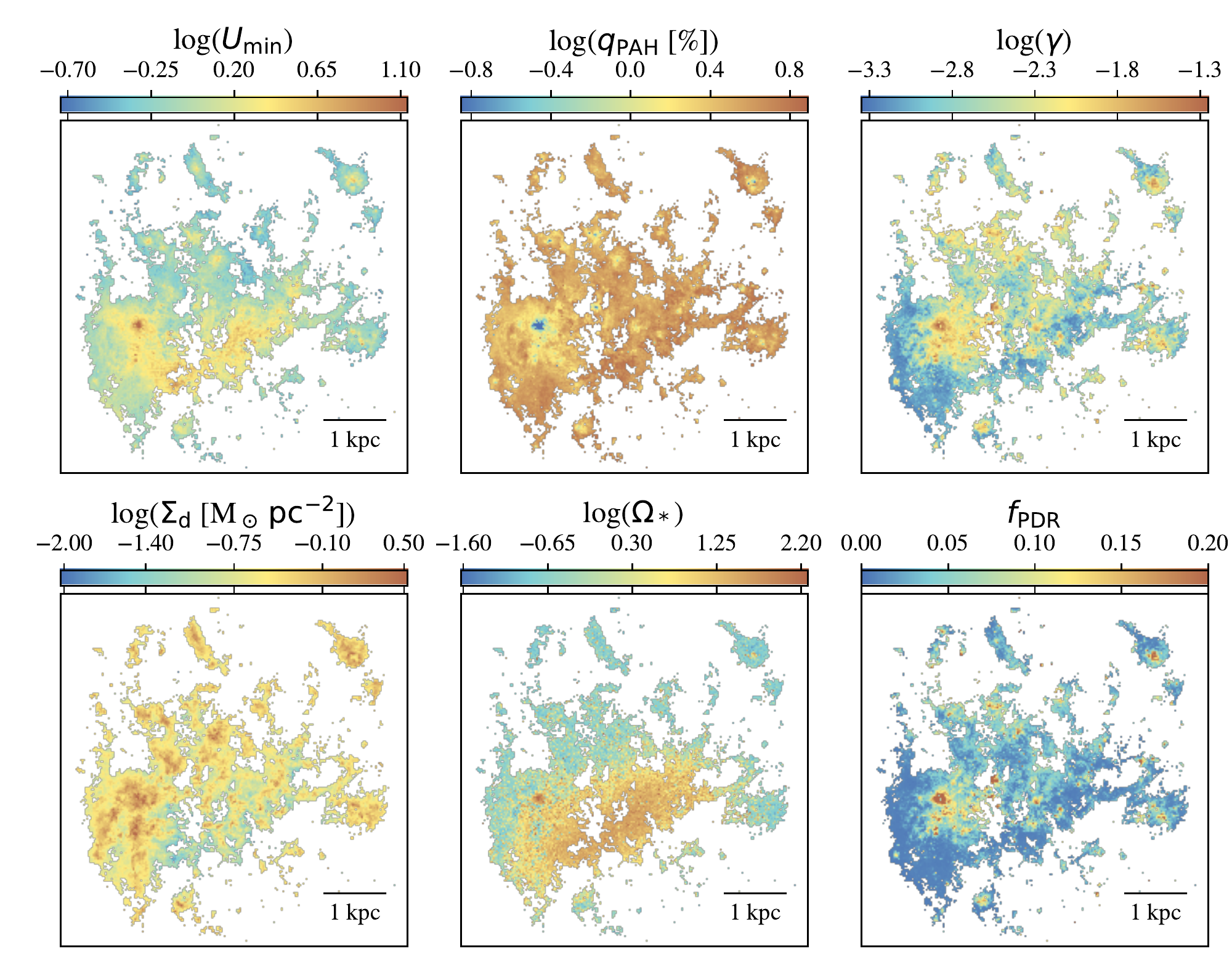}
    \caption{\small{Results of the fitting in the LMC for \umin, \qpah, $\gamma$, $\Sigma_{\rm d}$, and $\Omega_*$, and the derived \fpdr.}}
    \label{FigResultsLMC}
% trim={<left> <lower> <right> <upper>}
\end{figure*}

Figures \ref{FigResultsSMC} and \ref{FigResultsLMC} show the results from fitting the pixels of the SMC and the LMC maps, respectively, as well as the computed values of \ubar\ and \fpdr, as described in Section \ref{SecMethodo}. 
The overall distribution of $\Sigma_{\rm d}$ shifts towards higher values in the LMC than in the SMC, as expected because of the larger dust-to-gas ratio of the former \citep{Gordon14}.
A significant difference is also seen in the \qpah\ parameter, which is clearly higher in the LMC than in the SMC (a larger version of the \qpah\ map is shown in Figure \ref{FigMCsqPAHImg}). We find dust-mass-weighted average for each galaxy of 
\begin{align*}
    \langle q_{\rm PAH}^{\rm SMC} \rangle &= 1.0^{+0.3}_{-0.3}~\%, {~\rm and}\\
    \langle q_{\rm PAH}^{\rm LMC} \rangle &=  3.3^{+1.4}_{-1.3}~\%. 
\end{align*}
Both of these values fall within the range seen by \citet{Draine07} in the SINGS sample at the relevant metallicities.
By studying a sample of low-metallicity galaxies, \citet[][]{RemyRuyer15} found a power-law relation between the metallicity and the PAH fraction (see their Equation 5). If we apply this relation to the MCs metallicities \citep[SMC: 12+log(O/H) $\sim 8.1$; LMC: 12+log(O/H) $\sim 8.3$;][]{RusselDopita92}, we find:
\begin{align*}
0.69~\% &\leq q_{\rm PAH}^{\rm RR15}({\rm SMC}) \leq 3.47~\%, {~\rm and}\\
1.25~\% &\leq q_{\rm PAH}^{\rm RR15}({\rm LMC}) \leq 6.31~\%.
\end{align*}
Our average PAH fractions found in this work fall well within these ranges as well.

In both galaxies over most of the area, \umin\ $\sim$ \ubar. Thus we only show the fitting parameter \umin\ in Figures \ref{FigResultsSMC} and \ref{FigResultsLMC}. We calculate the dust-mass-weighted average of \ubar\ and find \avubar~=~1.2$^{+0.4}_{-0.4}$ in the SMC and \avubar~=~1.6$^{+1.6}_{-1.0}$ in the LMC. 
\citet[][]{Utomo19} recently studied the distributions of mass and temperature in four nearby galaxies, including the SMC and the LMC, using a single-temperature modified blackbody model. 
They find the distribution of dust mass as a function of radiation field intensity peaks at values of $U_{\rm peak} = 1.1$ in the SMC and $U_{\rm peak} = 1.8$ in the LMC, which are consistent with our results.

\citet{Sandstrom10} found a dust-mass-weighted PAH fraction of $\sim 0.6~\%$ in the SMC (compared to the 1.0~\% in this study). 
One possible explanation for our lower fraction is the limited coverage of the dust emission SED in \citet{Sandstrom10} compared to this paper: here, the longest wavelength is 500~$\mu$m while it is 160~$\mu$m in \citet[][]{Sandstrom10}. With the addition of the Herschel bands, we are able to better constrain the total dust mass and temperature, particularly in regions with colder dust, affecting the PAH fraction. In addition, the extent of the S$^3$MC {\em Spitzer} maps used in that paper did not allow as accurate a MW foreground removal as enabled by the expanded SAGE-SMC coverage.  This may have resulted in an oversubtraction of actual SMC emission in the mid-IR bands, decreasing the \citet{Sandstrom10} \qpah\ value.

In the LMC, \citet[][]{Paradis09} found an enhanced PAH fraction with respect to the large grain abundance in the stellar bar. Our results do not show a increased \qpah\ in this region, as other regions of the LMC show the same PAH fraction as in the optical bar. This difference could be the result of including the Herschel bands, and thereby more accurately measuring \qpah. It may also be the result of using a different dust model \citep[e.g. the \citet{DBP90} model in the paper by][]{Paradis09}, or a different stellar continuum model.

We find values of the global PDR fraction (i.e. the fraction of the dust luminosity produced by regions where $U \geq 10^2$): 
\begin{align*}
    \langle f_{\rm PDR}^{\rm SMC} \rangle &= 3.29^{+0.01}_{-0.02}~\%, {~\rm and}\\
    \langle f_{\rm PDR}^{\rm LMC} \rangle &= 9.17^{+0.04}_{-0.04}~\%.  
\end{align*}
These values are below the average values found in NGC 628 and NGC 6946, two nearby, resolved galaxies, by \citet[][$\sim 11~\%$]{Aniano12}. However, it is not surprising as the spatial scales in their study are coarser than for the MCs: due to limited resolution, the luminosity-weighted \fpdr\ is biased towards high-luminosity values. The resulting weighted-average is then more sensitive to these high-luminosity regions, due to the blending of signal. This is one of the key results of the recent study by \citet{Utomo19}.
We do find high \fpdr\ values in star forming regions, up to $\sim 60~\%$ in 30 Dor (LMC), and $\sim 50~\%$ in N66 (SMC).

The spatial distributions of $\Omega_*$ trace the regions with high stellar density \citep{Zaritsky02smc, Zaritsky04lmc}. It shows the old stellar spheroid population of the SMC, as well as the optical bar in the LMC.

\subsection{PAHs in the Magellanic Clouds}
\label{SecPAHinMC}
\begin{figure*}
    \centering
    \includegraphics[trim={0 1cm 0 2cm}, clip, width=0.92\textwidth]{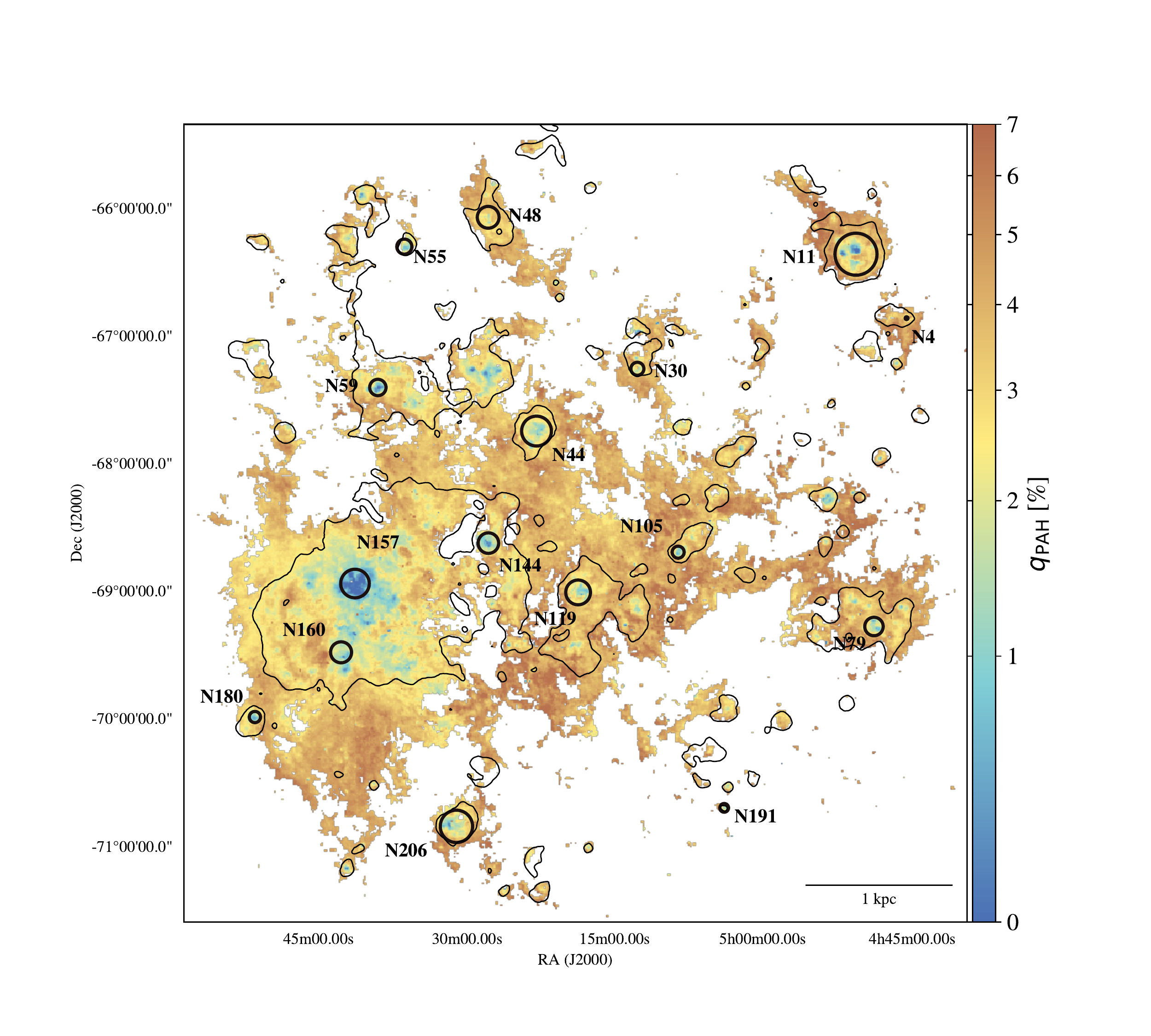}
    \includegraphics[trim={0 3cm 0 4.5cm},clip,width=0.83\textwidth]{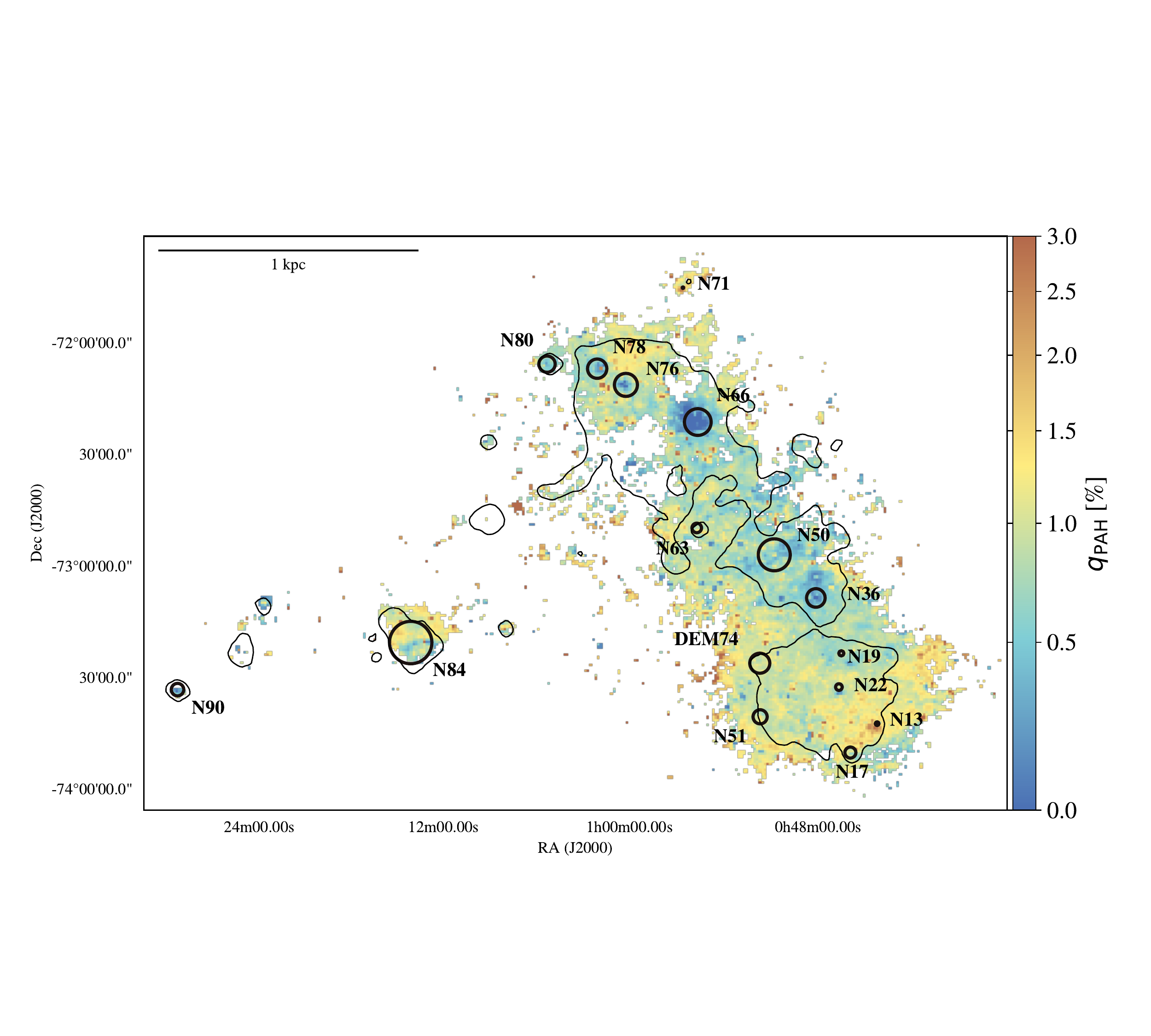}
    \caption{\small{Maps of the \qpah\ parameter in the LMC (top) and the SMC (bottom). We find dust-mass-averaged value of $\langle q_{\rm PAH}^{\rm SMC} \rangle=1.0~\%$, and
    $\langle q_{\rm PAH}^{\rm LMC} 
    \rangle=3.3~\%$. The contours are \halpha\ emission from the SHASSA survey at $\sim 1.5\times 10^{-16}~$erg/s/cm$^2$/arcsec$^2$. The labeled circles are \ion{H}{2} regions identified by \citet{Lopez14}. See Section~\ref{SecPAHinMC} for details.}}
    \label{FigMCsqPAHImg}
\end{figure*}

In Figure \ref{FigMCsqPAHImg}, we show the maps of the \qpah\ parameter in the LMC (top) and the SMC (bottom). The contours are H$\alpha$ emission from the SHASSA survey \citep[][]{Gaustad01} at level of $\sim 1.5\times 10^{-16}~$erg/s/cm$^2$/arcsec$^2$ ($\sim300$~dR; solid line; the reason for choosing this value is discussed later in this Section). The labeled circles are \ion{H}{2} regions as identified in \citet[][]{Lopez14}; the radii correspond to those given in their Table 1, and defined as the limit where they enclose 90~\% of the \halpha\ emission of the source. There are other known \hii\ regions in the Magellanic Clouds; however, identification of \hii\ regions and their boundaries is not straightforward, and we use only the ones in \citet{Lopez14} for the sake of homogeneity.
From these maps we draw several conclusions which we discuss in the following (sub)sections: 1) \qpah\ varies dramatically between the SMC and LMC, 2) \qpah\ shows variations \emph{within} each galaxy and 3) a primary driver for the variation in \qpah\ within each galaxy appears to be correlated with the presence of ionized gas as traced by H$\alpha$. 

To investigate the variations of \qpah\ within and between the galaxies, we make four separations with respect to the ISM gas phase.
In Table \ref{TabqPAH}, we report \avqpah\ in these four phases. There is no overlap between these values as we use only pixels that do not fall in two gas phase definitions. In short, we observe the following:
\begin{equation}
    q_{\rm PAH}^{\rm mol} \sim q_{\rm PAH}^{\rm diff.~neutr.} > q_{\rm PAH}^{\rm out-H\ II} > q_{\rm PAH}^{\rm H\ II}
\end{equation}
\noindent The implications of these observations for the PAH life-cycle will be discussed further in Section~\ref{SecPAHsgas}. We define the four phases as follows:
\begin{itemize}
    \item Ionized gas toward \hii\ regions.
    This is simply defined by the pixels falling within the radii of \hii\ regions from \citet[][]{Lopez14}.
    In the LMC, there is a clear drop in the dust-mass-weighted PAH fraction, and $\langle q_{\rm PAH}^{\rm H\ II} \rangle$ reaches only $\sim 1.8^{+1.1}_{-1.3}~\%$, i.e. slightly more than half of the galaxy average. In the SMC, the dust-mass-weighted PAH fraction reaches $\sim~0.8^{+0.3}_{-0.5}~\%$. We note that the harder radiation field in and near \hii\ regions, which is not taken into account in our fitting, would lead us to overestimate \qpah\ in these regions. The values found here are therefore conservative.
    \item Non-\hii\ region ionized gas.
    We distinguish the ionized gas inside and outside \hii\ regions, by selecting pixels whose \halpha\ surface brightness is above $I_{\rm H\alpha} \sim 1.5\times 10^{-16}~$erg/s/cm$^2$/arcsec$^2$ but not in identified \hii\ regions. Although this ionized gas is in a more diffuse phase than the gas in HII regions, we avoid identifying it as ``diffuse ionized gas (DIG)'' due to the specific ways that DIG is defined in nearby galaxies \citep[for reviews, see][ and references therein]{Mathis00,Haffner09}.
    We discuss this further in Section \ref{SecCaveats}.
    We find $\langle q_{\rm PAH}^{\rm out-H~{\small II}} \rangle = 2.9^{+1.1}_{-1.2}~\%$ in the LMC and $\langle q_{\rm PAH}^{\rm out-H~{\small II}} \rangle = 0.9^{+0.3}_{-0.3}~\%$ in the SMC.
    \item Molecular gas.
    We use \cogas\ maps (Section \ref{SecData}) to trace the molecular gas. We define this phase with every pixel above the 3-$\sigma$ detection threshold, and $I_{\rm H\alpha} \lesssim 1.5\times 10^{-16}~$erg/s/cm$^2$/arcsec$^2$ . 
    We find $\langle q_{\rm PAH}^{\rm mol.} \rangle \sim 4.3^{+1.3}_{-0.9}~\%$ in the LMC, and $\langle q_{\rm PAH}^{\rm mol.} \rangle \sim 1.1^{+0.1}_{-0.2}~\%$ in the SMC, similar to the values in the diffuse neutral medium.
    \item Diffuse neutral gas.
    \label{SubSubSecDNM}
    This is defined with the regions that fall into none of the above categories: with \halpha\ emission lower than $\sim 1.5\times 10^{-16}~$erg/s/cm$^2$/arcsec$^2$, which also means outside of an \hii\ region, and below the 3-$\sigma$ CO detection.
    We use the dust-mass-weighted PAH fraction in the diffuse neutral medium as a reference value for each cloud. We find $\langle q_{\rm PAH}^{\rm ref} \rangle = 1.1^{+0.2}_{-0.3}~\%$ in the SMC, and $\langle q_{\rm PAH}^{\rm ref} \rangle = 4.1^{+0.6}_{-0.8}~\%$ in the LMC.
\end{itemize}

\renewcommand{\arraystretch}{1.3}
\begin{deluxetable}{lll}
    \centering
    \caption{\avqpah\ (in \%) in different gas phases of each galaxy.}
    \tablehead{
    \colhead{} & \colhead{SMC} & \colhead{LMC}} 
    \startdata
    Global average & 1.0$^{+0.3}_{-0.3}$ & 3.3$^{+1.4}_{-1.3}$ \\
    \hline
    \hii\ regions & 0.8$^{+0.3}_{-0.5}$ & 1.8$^{+1.1}_{-1.3}$ \\
    Non-\hii\ region ionized gas\tablenotemark{1} & 0.9$^{+0.3}_{-0.3}$ & 2.9$^{+1.1}_{-1.2}$ \\
    Diffuse neutral\tablenotemark{2} & 1.1$^{+0.2}_{-0.3}$ & 4.1$^{+0.6}_{-0.8}$ \\
    Molecular gas\tablenotemark{3} & 1.1$^{+0.1}_{-0.2}$ & 4.3$^{+1.3}_{-0.9}$ \\
    \enddata
\tablenotetext{1}{All pixels above the lower limit in \halpha, excluding \hii\ regions.}
\tablenotetext{2}{All pixels below the CO detection, the limit in \halpha, and outside of \hii\ regions.}
\tablenotetext{3}{All pixels with CO detection not overlapping with pixels with $I_{\rm H\alpha} \gtrsim 1.5\times 10^{-16}~$erg/s/cm$^2$/arcsec$^2$.}
\label{TabqPAH}
\end{deluxetable}

\subsection{\qpah\ in \hii\ Regions}
\label{SecPAHinHII}
The \ion{H}{2} regions, as identified in \citet[][]{Lopez14}, appear as minima in the \qpah\ map, and suggest that PAHs are destroyed inside of \hii\ regions.
In Figure \ref{FigMCsqPAHImg}, we can see that most \hii\ regions, marked by the black labeled circles, are indeed low in PAHs, with respect to the abundance in the other parts of each galaxy. 

\begin{figure}
     \centering
     \includegraphics[width=0.48\textwidth]{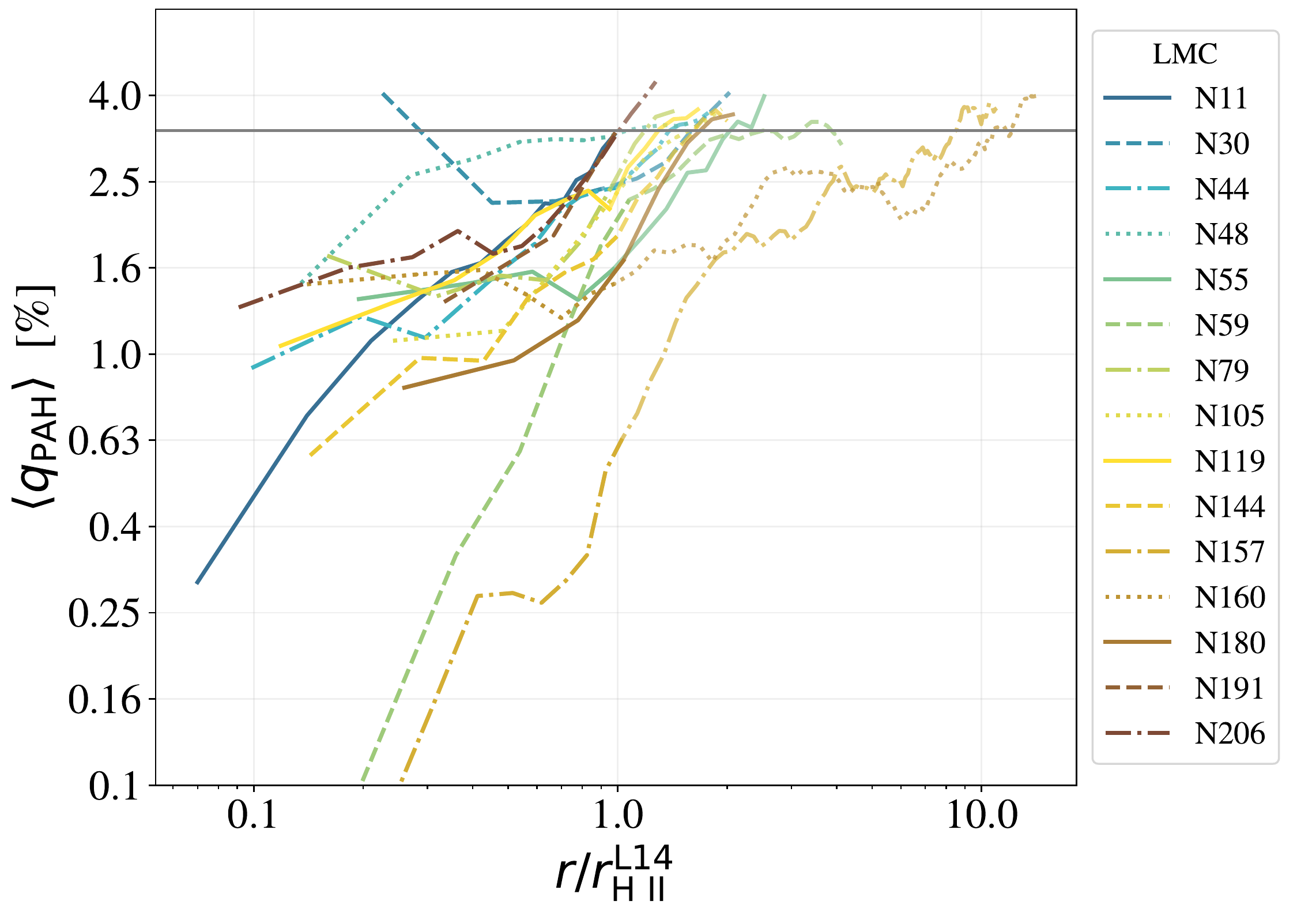}
     \includegraphics[width=0.48\textwidth]{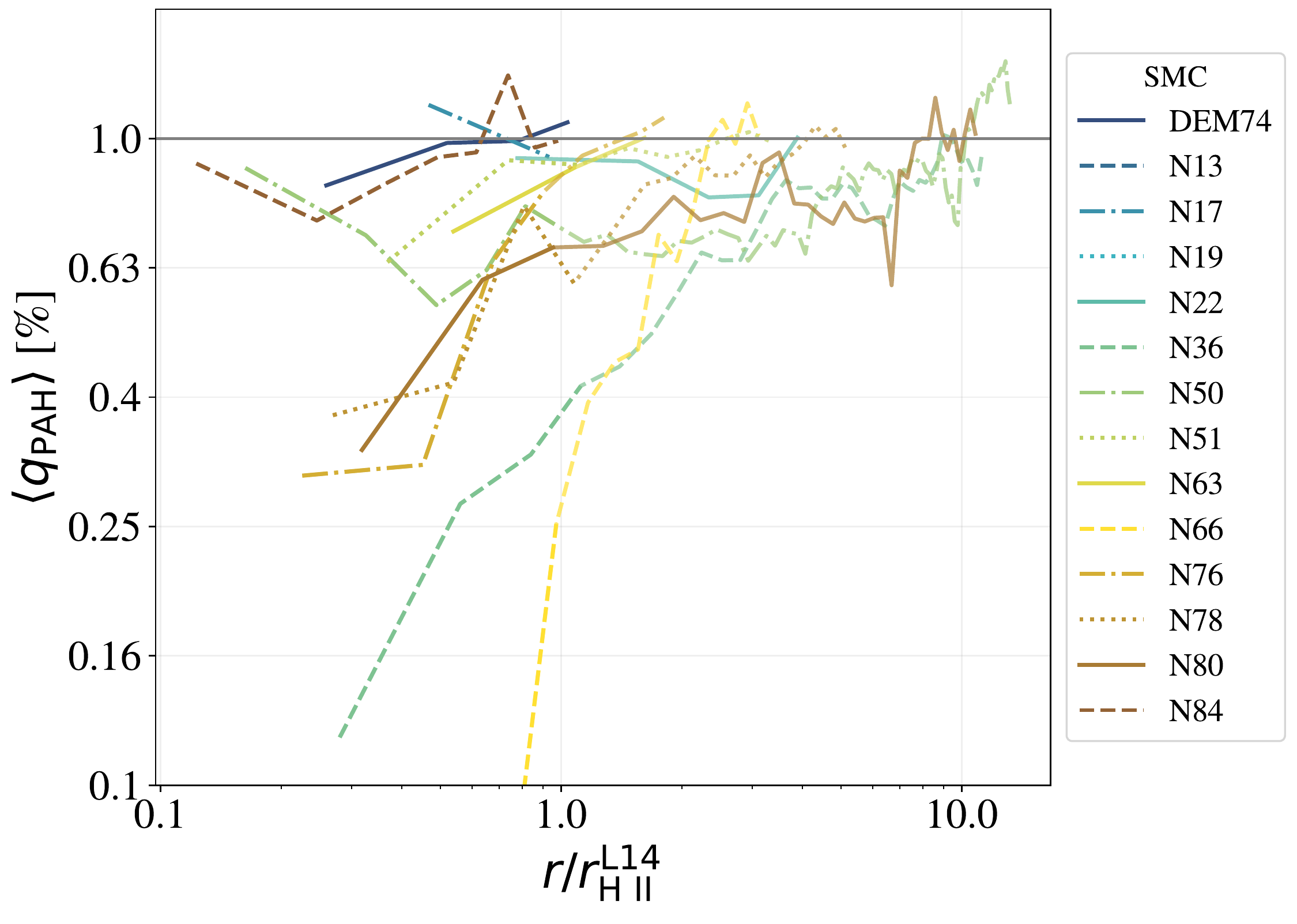}
     \caption{\small{Radial profiles of \avqpah\ in the LMC (top) and the SMC (bottom), from the centers of each \hii\ region, normalized by the radius given by \citet[][]{Lopez14}. The horizontal grey lines mark the average for each galaxy. The maximum radii are chosen to be slightly higher than the radii of the spheres of influence, in order to show its full extent.}}
     \label{FigMCProfiles}
 \end{figure}

We are interested in understanding how \hii\ regions affect the PAH fraction. 
In Figure \ref{FigMCProfiles}, we show the radial profiles of \avqpah\ for each \hii\ region in the SMC (bottom) and the LMC (top). 
We see that \avqpah\ drops to very low values only in a handful of cases (notably, 0~\% in 30 Doradus). This is expected, as the diffuse neutral gas projected along the line-of-sight contaminates our measurement of the \qpah\ inside the \hii\ region. The dust emission along the line-of-sight reflects the full column through the ISM of the galaxy, not just the \hii\ region, so we do not expect the observed \qpah\ to be 0~\%. In the particular case of 30 Doradus, a result consistent with 0~\% PAH fraction could mean that the actual \hii\ region dominates the entire line of sight through the LMC.

\begin{figure}
     \centering
     \includegraphics[width=0.49\textwidth]{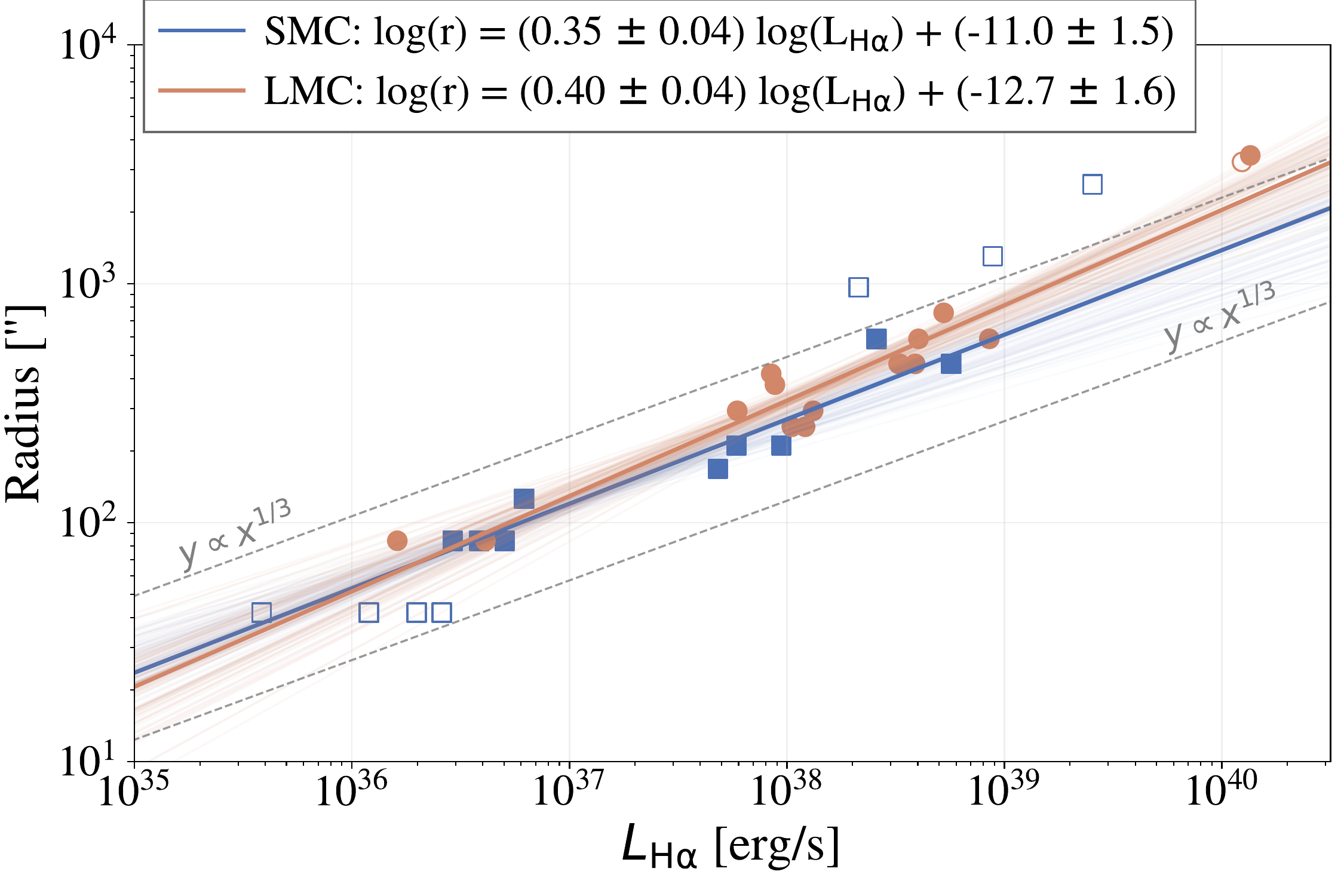}
     \caption{\small{Correlations between the total \halpha\ luminosity and the radius at which \avqpah\ returns to the global average from the center of each \hii\ regions. The slope found in both cases is close to that expected for the expansion of a typical Str\"omgren sphere. We add $y=x^{1/3}$ slope for comparison (grey dashed lines). (The empty symbols mark region where the calculated radius exceeds five times the one given in \citet[][]{Lopez14}, and are not used to fit the line. The light lines are results from bootstrapping for uncertainties on the coefficients.)}}
     \label{FigRadiiInfluence}
 \end{figure}

For each \hii\ region, we measure the radius at which \avqpah\ goes back to the global average (horizontal grey line in Figure \ref{FigMCProfiles}).
In this Figure, we see that \avqpah\ in each \hii\ region goes back to the galaxy average at different radii. 
We treat these radii as the ``spheres of influence'' of each \hii\ region on the surrounding PAH fraction. In Figure \ref{FigRadiiInfluence}, we plot these radii against the total \halpha\ luminosity from within that radius. 
We find that these radii correlate very well with \halpha\ luminosity with a power-law coefficient between 0.35 and 0.4, in both galaxies. If the \halpha\ surface brightness were constant, growing the radius would lead to a dependence of $L_{\rm H\alpha}$ in $r^2$. We see in Figure~\ref{FigRadiiInfluence} that is not the case.
It rather suggests that the sphere of influence on the PAH fraction of the ionizing stars within an \hii\ region scales as one would expect for the Str\"omgren sphere \citep[][]{Stromgren39}. In this particular case, assuming a constant gas density the radius of the Str\"omgren sphere grows with the ionizing photon production rate (as does the H$\alpha$ luminosity) with a power-law coefficient of 1/3. If PAHs are destroyed in ionized gas by photo-destruction or sputtering, then one would expect the region with a deficit of PAHs relative to the galaxy average to grow as the size of the Str\"omgren sphere.
In \citet[][]{Binder18}, the authors found a relation between the population of stars inside star forming regions of the MW, and the PAH fraction. They show that a single O6 star is less effective at destroying PAHs than a population of stars that extends to O2/O3, and Wolf-Rayet stars. 
\citet[][]{Glatzle19} showed that the growth of \hii\ regions is related to the dust content within, including PAH abundance, by impacting the ionization fronts.
A more detailed study of \hii\ regions would be interesting to possibly link the initial \avqpah\ value at small radius in Figure~\ref{FigMCProfiles}, the properties of the ionizing star(s) within and the expansion of \hii\ regions.

\subsection{\qpah\ and \halpha}
\begin{figure}
    \centering
    \includegraphics[width=0.49\textwidth]{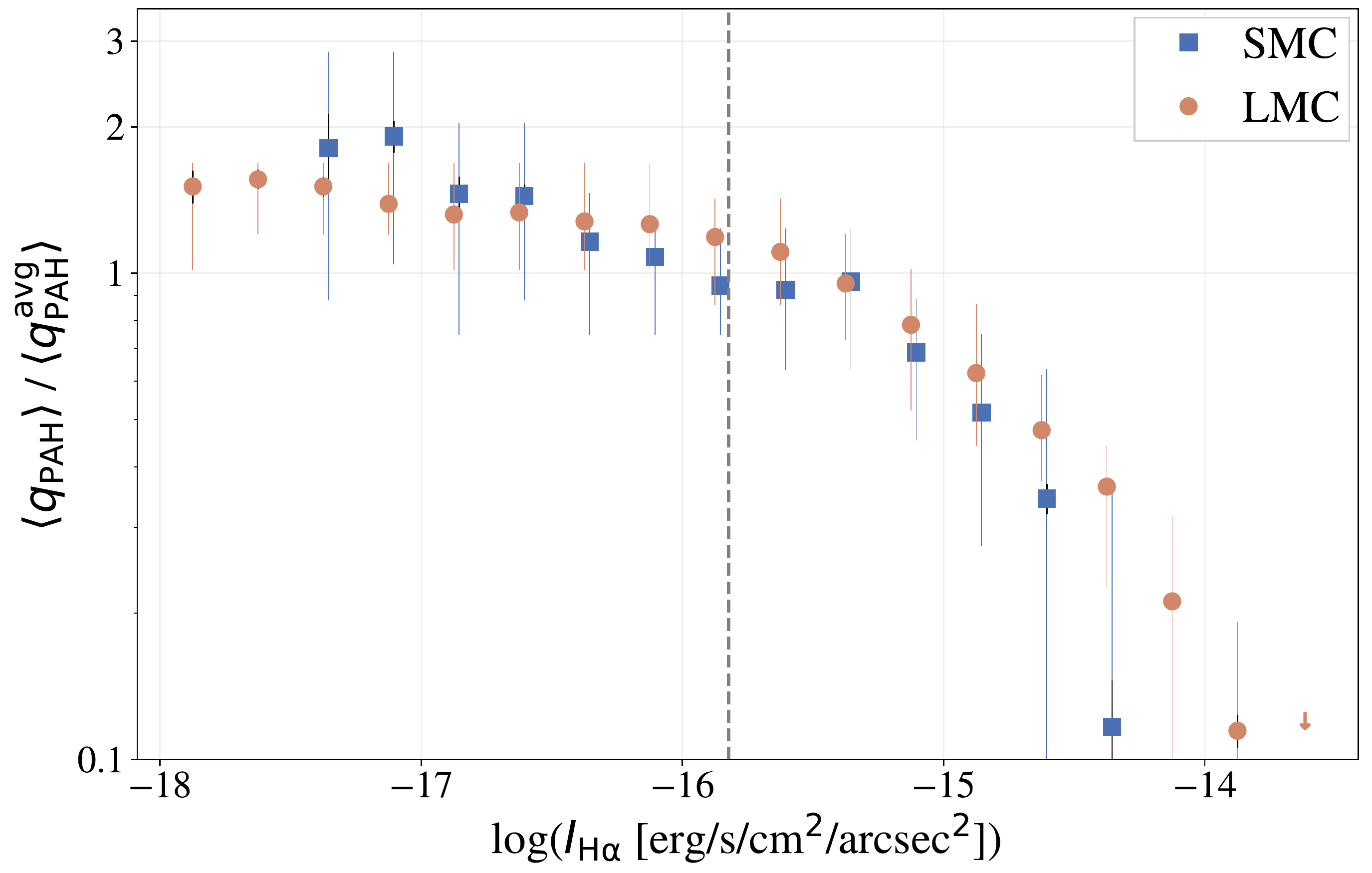}
    \caption{\small{Correlation between \avqpah\ and  \halpha\ surface brightness, for the LMC (orange circles) and the SMC (blue squares). The PAH fraction is normalized to the global average in each galaxy. Arrows indicate that the value falls below \avqpah\ = 0.1. The vertical dashed gray line marks the limit between the diffuse neutral and non-\hii\ region ionized gas.}}
    \label{FigMCsGasCorr}
\end{figure}

Even outside of the \hii\ regions, we find a clear relation between increasing \halpha\ emission and decreasing \avqpah. 
Figure \ref{FigMCsGasCorr} shows \avqpah\ binned as a function of H$\alpha$ surface brightness for the SMC (blue squares) and the LMC (orange circles). 
Both correlations tend to reach $\langle q_{\rm PAH} \rangle \sim 0~\%$ at high $I_{{\rm H}\alpha}$ values, independently of the average value of each galaxy. This agrees with the scenario where PAHs are destroyed in \hii\ regions.
In Figure \ref{FigMCsGasCorr}, we identify the value $I_{{\rm H}\alpha} \sim 1.5\times 10^{-16}~$erg/s/cm$^2$/arcsec$^2$ (or $\sim 300~$dR) as the \halpha\ emission level at which \avqpah\ starts to rapidly decrease in the LMC (dashed gray line). We use this value to plot the solid contours on Figure \ref{FigMCsqPAHImg}.
This marks a limit where the PAH fraction drops more steeply with \halpha\ surface brightness. We use this value as a separation between the diffuse neutral gas and ionized gas.
% The existence of that lower limit is interesting as it hints at a fraction of ionized gas along the line of sight, at which it contributes significantly to the observed PAH fraction.
Figure~\ref{FigMCsGasCorr} shows that the PAH fraction decreases even at low values of \halpha\ surface brightness, but that it drops rapidly only after it reaches the vertical dashed line ($\sim 1.5\times 10^{-16}~$erg/s/cm$^2$/arcsec$^2$, or ~300~dR). Before that value, the PAH fraction decreases only mildly. To first order the \halpha\ emission traces the surface density of ionized gas. Given that the distribution of \ion{H}{1} is fairly flat across both galaxies \citep[e.g.][]{Stanimirovic00}, it is possible that the turn-over at 300~dR occurs at a point where ionized gas starts to make a large contribution to the total surface density.
% Rather than the absolute \halpha\ emission itself, we consider that it is possibly a fraction of ionized gas in the line of sight that is responsible for the steep drop at 300~dR.}
The limit marked by the vertical dashed line is lower than the \halpha\ surface brightness in \hii\ regions, implying that PAHs may undergo destruction even in ionized gas outside of \hii\ regions.\footnote{Assuming an electron, and proton, density n$_{\rm e} \sim n_{\rm p} = 0.5~{\rm cm^{-3}}$, a gas temperature T~$=~5000~$K, and a flat \ion{H}{1} surface density in the LMC $\Sigma_{\rm H I} = 20\times 10^{20}~{\rm cm^{-2}}$, our limit in \halpha\ surface brightness corresponds to a fraction of ionized gas $\sim 12~\%$, or an ionized zone depth $\sim 150~$pc.}

A decrease of PAH fraction in diffuse ionized gas has been suggested by previous works in the Milky Way \citet[][]{Dobler09} and \citet{Dong11}. In \citet[][]{Dong11}, the authors studied the \halpha-to-free-free emission in the MW diffuse ionized gas, and found that this ratio corresponds to a lower temperature than they were able to produce with their model using the MW diffuse ISM PAH abundance. With a model consisting of ionized gas, recombining gas in the process of cooling, and cool neutral gas, they manage to reproduce the observations by lowering (by a factor of $\sim 3$) the PAH fraction in photo-ionized regions, with respect to that of the global average in the ISM. Our results, constrained from observed SEDs, agree with a scenario where the PAH fraction is lower in ionized gas.

\subsection{\qpah\ and the radiation field}
\label{SecUbarqPAH}
\begin{figure}
    \centering
    \includegraphics[width=0.49\textwidth]{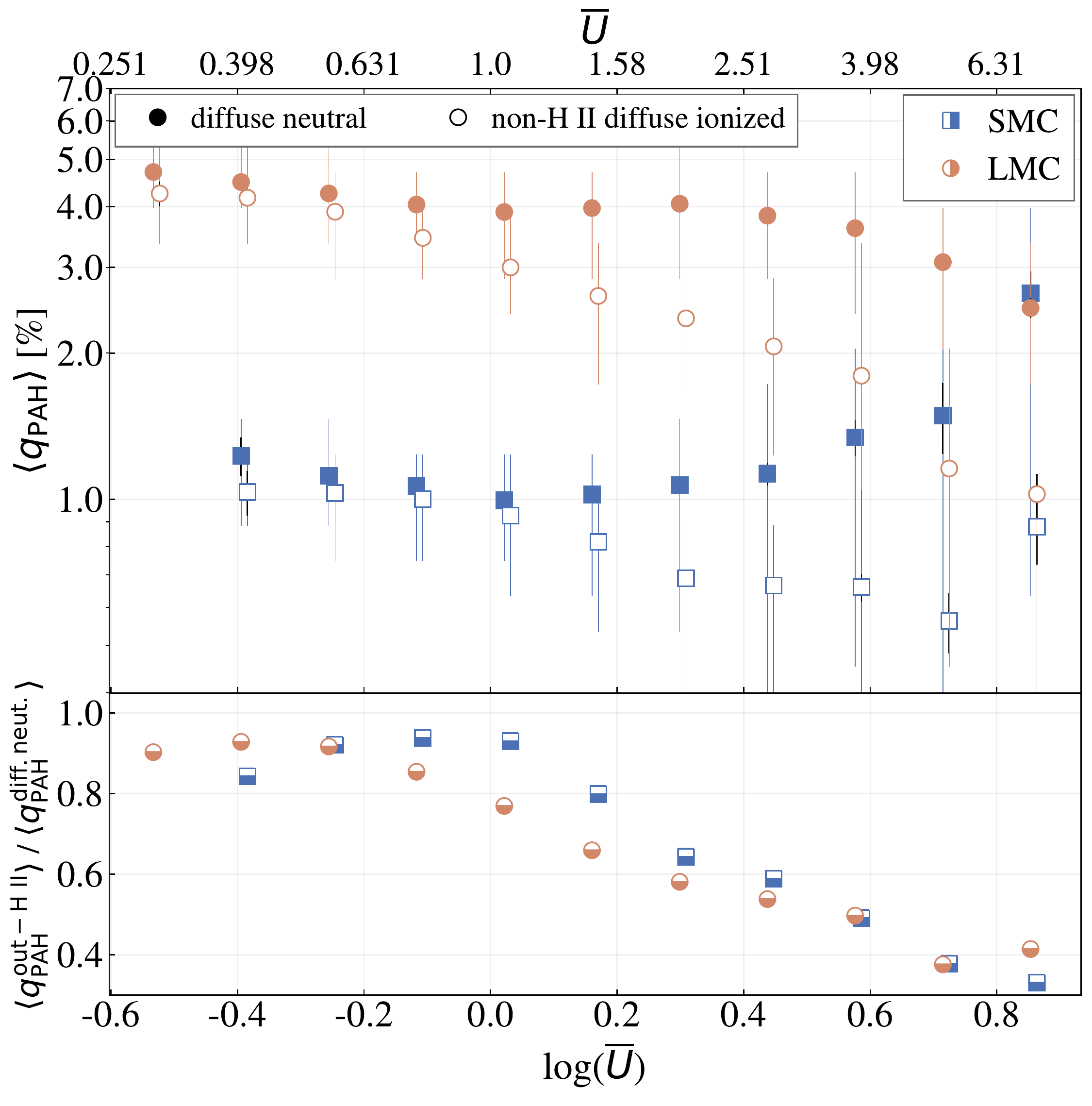}
    \caption{\small{{\it Top:} Variations of \avqpah\ as a function of \ubar\ derived from the fit, for the SMC (blue), and the LMC (orange). Pixels that fall above the limit in \halpha\ emission, and outside of \hii\ regions, are marked with empty symbols. Pixels that fall in the diffuse neutral medium, i.e. below 300~dR and below the CO detection threshold, are marked with filled symbols. {\it Bottom:} Fraction of \avqpah\ in the ionized medium with respect to \avqpah\ in the neutral medium.}}
    \label{FigUbarqPAH}
\end{figure}

A high intensity of the radiation field is often quoted as a cause for enhanced destruction of PAHs. Here, we can test that scenario by looking at \avqpah\ as a function of \umin, or \ubar. In our work, we only adjust for the radiation field intensity, and not the hardness. 
In the top panel of Figure \ref{FigUbarqPAH}, we show the variations of \avqpah\ in bins of \ubar, in the diffuse neutral medium (filled symbols), and the non-\hii\ region ionized gas (above the limit in \halpha\ emission and outside of \hii\ regions; empty symbols). We choose to compare the diffuse neutral medium and the non-\hii\ region ionized gas because ionized-gas related destruction processes should not occur in the former. 
In the diffuse neutral medium of the SMC, \avqpah seems to increase slightly with \ubar although with increasing uncertainty. This uptick in \avqpah\ may be the result of an overestimation due to the fact that we do not take into account a change in radiation field hardness (see Section~\ref{SecRF}) as \ubar\ increases. It is possible that in low-metallicity environments, changes in the radiation field hardness are more important, hence the observed increase of \avqpah\ in the SMC.
In the diffuse neutral medium in the LMC however, we can notice a decrease of \avqpah\ at \ubar\ $\sim3$. Given the scatter, there are only minor variations of \avqpah\ up to log(\ubar)~=~0.5 (\ubar\ $\sim 3$). At this point, \avqpah\ decreases noticeably. Below this, the intensity of the radiation field does not seem to affect the PAH fraction. As expected, in the ionized gas outside of \hii\ regions, there seems to be a decrease of \avqpah\ with \ubar\ even for \ubar $ < 1$. This is not surprising because in defining this phase, we selected the pixels where \avqpah\ decreases in Figure \ref{FigMCsGasCorr}.
When we control for the intensity of the radiation field, we can see that the gas phase has an impact of the PAH fraction. For an identical \ubar, the PAH fraction does not decrease as quickly, whether it is in the diffuse neutral or the non-\hii\ region ionized gas.
We also point out that, even at the lowest \ubar, \avqpah\ in the ionized medium never reaches that of the neutral medium. 

The bottom panel of Figure \ref{FigUbarqPAH} shows the ratio of PAH fraction in different gas phases $\langle q_{\rm PAH}^{\rm out-H~II} \rangle / \langle q_{\rm PAH}^{\rm diff. neut.} \rangle$ (same color code than the top panel). It is interesting to notice the increasing offset between the PAH fraction in the ionized gas and the neutral medium as \ubar\ increases and that this offset is the same in both galaxies. 
We discuss these results in the context of PAH destruction in Section \ref{SecPAHsgas}.

\section{Discussion}
\label{SecDiscussion}

\subsection{Dependence of \qpah\ on metallicity}
The difference between the global averages $\langle q_{\rm PAH}^{\rm SMC} \rangle$ and $\langle q_{\rm PAH}^{\rm LMC} \rangle$ is in good agreement with the power-law dependence of \qpah\ with metallicity found by \citet[][]{RemyRuyer15}\footnote{Their study deals with integrated SED fitting. Their \qpah\ parameter is then similar to a luminosity-mass weighted \avqpah\ in our study.}. Our values also sit well within the scatter of the \citet{Draine07} results, where a steep drop of \qpah\ occurs at a metallicity threshold of 12+log(O/H) $\sim 8.1$. 

The degree to which our results agree with either the threshold or power-law dependence, however, depends on the gas phase. In the MW, studies have found diffuse neutral medium  \qpah\ value of $\sim 4.6~\%$ \citep{LD01,WD01}. For a fair comparison, this should be compared to our selection of the diffuse neutral medium in the LMC and SMC listed in Table~\ref{TabqPAH}. In this case, the power-law boundaries found by \citet[][]{RemyRuyer15} underestimate the PAH fraction in the LMC. Indeed, the LMC's diffuse neutral gas \qpah\ is very similar to the Milky Way's, while the SMC falls a factor of $\sim4$ lower. If we treat the diffuse neutral gas as a reference value, where most destruction processes are not operating, our results strengthen the interpretation that there is a threshold in metallicity where \qpah\ decreases significantly, and that is lies between SMC and LMC metallicities.

\subsection{Insights into the life-cycle of PAHs from comparison of \avqpah\ in different phases}
\label{SecPAHsgas}
We use the separation of the gas phases in the MCs and their associated PAH fraction to investigate the life-cycle of PAHs in the ISM. We first go over some of the possible scenarios for PAH formation and how they might affect \avqpah\ in each phase. 

Theoretical studies have found that the atmospheres of post-AGB stars could be a favored site for the formation of PAHs \citep[][]{Cherchneff92, Sloan07}. 
PAHs produced by AGB stars should be primarily input into the diffuse ISM, since the old stellar distribution is not closely related to the current location of star-forming regions or dense gas \citep[e.g. see][for a comparison between the SMC old stellar distribution and ISM]{Sandstrom10}. If PAHs are primarily formed by AGB stars and no additional destruction or production mechanisms are operating in the diffuse neutral gas, one could argue that the $\langle q_{\rm PAH}^{\rm ref} \rangle$ we have identified should reflect the efficiency of PAH formation by AGB stars. \citet{Boyer11} and \citet{Boyer12} studied the stellar population of both MCs, and found that the carbon-rich AGB population have similar dust production rates in the SMC and in the LMC, despite their different metallicities. Although other evolved stars such as red supergiants seem to be more effective at producing dust in the LMC, they do not play a major role in the overall dust production by evolved stars \citep[e.g.][]{Boyer11}. 
Therefore, without some additional destruction mechanism in the diffuse neutral phase, we cannot explain the different \avqpahref\ between the SMC and the LMC solely with more effective formation of PAHs in AGB atmospheres in the LMC\footnote{If the PAH-like dust production is equal in both clouds, a higher production of non-PAH dust in the SMC would lead to a lower PAH \emph{fraction}. Based on the same works previously quoted, there is no indication than the SMC stellar population is more effective at producing non-PAH dust than in the LMC, and we disfavor this possibility.}.

Another scenario for PAH formation suggested by theoretical studies \citep[e.g.][]{Jones96} is shattering of large carbon grains leading to the production of smaller carbon grain fragments. 
\citet{Jones96} argued that the shattering of carbon grains can happen at velocity as low as 1~km~s$^{-1}$, and is the prevalent process affecting dust grains at shock velocities $\geqslant 100~$km~s$^{-1}$. 
Their study focused on the grain-grain interaction in the so-called warm intercloud medium, similar to our definition of diffuse neutral medium. There, they found that the redistribution of carbon-rich dust mass from large grains into small grains, and even PAH-like fragments, is significant. In \citet{Slavin15}, the authors found that the redistribution of dust mass in smaller grains is important in supernova remnants shocks up to $\sim 200~$km~s$^{-1}$.
A critical question, however, is whether the hot shocked gas is able to quickly sputter and destroy any PAH fragments that are created by shattering. Theoretical calculations by \citet{Micelotta10shocks} suggest that PAHs are quickly destroyed by collisions with energetic particles at shock velocities higher than 100~km~s$^{-1}$. This work leads to the conclusion that shattering of dust grains, in $<100$~km~s$^{-1}$ shocks could be a significant source of PAH-like grains. Because the shattering process is directly linked to the available carbon-rich dust mass, one would expect a higher abundance of small grain fragments if there is a higher dust-to-gas ratio. Given the observed dust-to-gas ratio in the Magellanic Clouds \citep[$\sim 0.003$, $\sim 0.0008$ in the LMC and SMC, respectively;][]{Leroy11,Gordon14,RomanDuval14}, this would be consistent with our finding of higher \avqpahref\ in the LMC, compared to that of the SMC. In the former, the higher \avqpahref, assuming destruction of PAHs is not significant in the diffuse neutral medium, could be the result of more efficient shattering of large carbonaceous grains.

Another hypothesis for PAHs formation is growth in the molecular phase of the ISM \citep[e.g.][]{Sandstrom10, Sandstrom12, Zhukovska16}. In this scenario, the lower \avqpah\ in the SMC would be related to less efficient growth processes in the molecular gas, possibly due to the lower abundance of metals to accrete onto existing grains to form PAHs. We do see that the $\langle q_{\rm PAH}^{\rm mol.} \rangle$ is lower in the SMC than in the LMC. Previous work by \citet{Sandstrom10} found enhanced \qpah\ in dense regions of the SMC compared to the diffuse gas phases, which was interpreted as evidence for growth in the molecular gas and destruction operating in the diffuse ISM.  We do find higher PAH fraction in the molecular gas phase of each galaxy with respect to the global averages,  (see Table \ref{TabqPAH}; LMC: $4.3~\%$; SMC: $1.1~\%$), but the \avqpah\ is similar in the molecular and diffuse neutral phases of each galaxy.
The observation of the same \avqpah\ in the molecular and diffuse neutral gas is consistent either with formation of PAHs in the diffuse ISM and incorporation into molecular clouds or its inverse, assuming no destruction processes operate differentially in these two phases.
In the same molecular gas, accretion and coagulation of small grains onto big grains \citep[e.g.][]{Stepnik03, Kohler12} could also lead to a decrease in the observed PAH fraction. In our case, observational evidence for that mechanism would be lower PAH fraction in the molecular regions. Since we do not see such variation, we do not favor this possibility.

Observational studies have shown that the PAH size distribution and properties are sensitive to the local radiation field and gas ionization. This is seen in the vicinity of Milky Way \ion{H}{2} regions and photo-dissociation regions \citep[e.g.][]{Berne07, Compiegne07, Arab12, Peeters17} as well as in nearby galaxies \citep[][]{Gordon08,Paradis11, Relano16}.
PAH destruction can be accomplished in several ways, mediated by interaction with energetic photons, sputtering by particles in hot gas, or chemical reactions. 
A long-standing hypothesis to explain the PAH deficit at low metallicity is that PAHs are more readily destroyed in such conditions \citep[e.g.][]{Madden06, Galliano08}. 
A possible scenario for enhanced destruction of PAHs at low metallicity is photo-destruction by the radiation field, either due to increased intensity or hardness. There are a number of observational studies that have found correlations that agree with such a scenario. For example, \citet{Madden06} and \citet[][]{Gordon08} found that the PAH features disappear as the radiation field hardness increases in \hii\ regions (measured from mid-IR neon and sulfur line ratios). 
Theoretical studies have also shown that PAHs are subject to sputtering and fragmentation in ionized gas \citep[][]{Micelotta10gas, Bocchio12} because of electronic- and/or atomic interactions. There, projectile particles can reach a high velocity with respect to that of the grains, leading to catastrophic collisions.
In general, the regions of the galaxy with the hardest radiation fields will also be those where ionized gas exists and the overall intensity of the radiation field is higher. Therefore, to distinguish between the potential destruction mechanisms, we need to attempt to separate the effects of these quantities. We do so using H$\alpha$ emission as a tracer of ionized gas, and our determination of \ubar\ as a tracer of the intensity of the radiation field, as presented in Section \ref{SecUbarqPAH} and Figure \ref{FigUbarqPAH}. We do not have a direct tracer for the hardness of the radiation field covering the full extent of the galaxies. 
We note the similar trends of the ratios $\langle q_{\rm PAH}^{\rm out-H~II} \rangle / \langle q_{\rm PAH}^{\rm diff. neut.} \rangle$ of the SMC (blue) and the LMC (orange), in the bottom panel of Figure \ref{FigUbarqPAH}. They suggest that metallicity does not have an impact on the relative efficiencies of the destruction processes between the diffuse neutral medium and the ionized gas outside of \hii\ regions (in the same bin of radiation field). We do not test, however, the impact of metallicity on the global amount of each gas phase in a galaxy, which might have an impact on the overall destruction of PAHs.

PAHs in a more intense radiation field could be more easily destroyed because they are more fragile, due to their ionization state. Ionized PAHs are less stable, and more prone to losing, for example, H atoms \citep{Montillaud13}.
\citet[][]{WD01charge} showed that the charge of small carbonaceous grains varies depending on a parameter G$\sqrt{\rm T}$/n$_{\rm e}$, where G is the radiation field intensity, T the temperature of the gas, and n$_{\rm e}$ the electron density.
To test whether PAH ionization may lead to higher destruction rates, the different behavior of the neutral and ionized gas phases at fixed \ubar\ from Figure~\ref{FigUbarqPAH} is of interest.
In a bin of \ubar, we control for the radiation field intensity, and it is the same in both the diffuse neutral and the non-\hii\ diffuse ionized gas. \citet[][]{WD01charge} showed that the gas temperature does not significantly affect the ionization state of PAHs. The only parameter remaining then is $n_e$.  
If the electron density were lower in the non-\hii\ ionized gas, compared to the diffuse neutral gas, then PAHs could be more ionized in this medium, facilitating their destruction. While this seems counter-intuitive, given that the non-\hii\ region ionized gas is defined by its ionization state, the difference in overall density between the ionized and neutral phases and the low, but non-negligible, fractional ionization in the neutral medium could lead to the situation where n$_e$ is lower for the ionized gas.
\citet{LD01} showed that PAHs larger than 7~\AA\ are more ionized in the Milky Way's \emph{warm} neutral medium than in the warm ionized medium. In that case, the lower $\langle q_{\rm PAH}^{\rm out-H~II} \rangle$ compared to that of the diffuse neutral medium could not be explained by an easier destruction of PAHs because of their ionization.

Based on the comparison between the gas phases in the two galaxies, we could explain the offset in \avqpah\ between the LMC and SMC with: (i) fragmentation of large grains in the diffuse medium leading to more PAHs in the LMC due to its higher dust content, (ii) formation of PAHs in the molecular ISM and their injection in the diffuse gas, assuming there is no preferential destruction in either molecular or diffuse gas. The formation of PAHs in AGB stars cannot explain the difference in PAH fraction between the SMC and the LMC. The destruction mechanisms (e.g. photo-destruction by the radiation field), suggested to be more effective in lower metallicity galaxies, do not differ significantly between the SMC and the LMC in this work (see bottom panel of Figure~\ref{FigUbarqPAH}).

\subsection{Caveats}
\label{SecCaveats}
\subsubsection{Impact of the modeled radiation field}
\label{SecRF}
The \citet{DL07} model uses the radiation field described by \citet{Mathis83}, for the Milky Way at the galactocentric distance $D_{\rm G} = 10~$kpc. This limits the possible variations allowed in our fitting. Specifically, the hardness of the radiation field is constant. However, it is expected that the relative proportion of UV vs optical photons will not be the same in \ion{H}{2} regions where the ionizing stars produce more UV photons. Some studies have made adjustments to the radiation field to address this issue \citep[e.g.][]{Galliano05, Salgado16}. In M33, \citet[][]{Relano16} studied the dust content of \ion{H}{2} regions, and specifically changed the radiation field to one with more UV photons. They showed that this leads to lower PAH fraction, by a factor up to 3. 
\citet{Paradis11} studied the impact of the radiation field on dust emission fitting in the LMC, by adding a 4~Myr stellar population to the \citet{Mathis83} radiation field. As expected, they found that, in \ion{H}{2} regions, using a harder radiation field leads to a decrease in the PAH abundance estimation. This is linked to the PAH being more sensitive to the UV-visible part of the incident radiation field \citep{Li02}. Harder radiation, i.e. more energetic photons, would enhance their MIR emission. In our study, this would make the offset between the SMC and the LMC \citep[low-metallicity stars produce more UV photons; e.g.][]{Eldridge08}, and the difference between diffuse-to-ionized PAH fraction, even more dramatic \citep[this was also shown by][in the LMC]{Paradis11}.

\subsubsection{Diffuse ionized gas}
The definition of ionized gas in the MW \citep[called ``warm ionized gas'', WIM; e.g.][]{Reynolds84, Madsen06}, and in extra-galactic studies \citep[called ``diffuse ionized gas'', DIG; e.g.][]{Zurita00} is a delicate subject \citep[see also][]{McKee90, Mathis00,Haffner09}. 
Given the difficulty defining ``diffuse ionized gas'', we decided not to make assumptions on the exact definition of the ionized medium outside of \hii\ regions.
A finer description of the local conditions of the ionized gas would require measurements of the electron/proton density, incident radiation field, and gas temperature.

\subsubsection{Metallicity variations across the Magellanic Clouds}
In this study, we assume the metallicity to be constant across each galaxy. 
If there were metallicity variations, there should also be variations in \qpah. If the metallicity variations are not correlated with ISM phase, we would only expect to see enhanced scatter. If they were correlated with ISM phase, the results would likely be different. We do not see a good reason to assume any metallicity variations would be ISM phase correlated.

\section{Conclusions}
We fit the dust emission SED in the Small and Large Magellanic Clouds, using photometry from \spitzer\ and \herschel\ (3.6 to 500~$\mu$m), with the dust emission model from \citet{DL07}. We provide maps of each fitting parameter at a 42'' pixel size, i.e. $\sim 10~$pc in the LMC and $\sim 12~$pc in the SMC.

We especially focus on the spatial distribution of the PAH fraction \qpah, the fraction of dust mass in grains with less than 10$^3$ carbon atoms. 
We find a global dust-mass-weighted PAH fraction $\langle q_{\rm PAH}^{\rm SMC} \rangle = 1.0~\%$ and $\langle q_{\rm PAH}^{\rm LMC} \rangle = 3.3~\%$, both lower than the Milky Way diffuse neutral medium value (4.6~\%). 

We measure the PAH fraction in different gas phases, distinguished by \cogas\ and \halpha\ emission (Table~\ref{TabqPAH}) or the lack thereof. We use the diffuse neutral medium ($I_{\rm H\alpha} \lesssim 1.5\times 10^{-16}~$erg/s/cm$^2$/arcsec$^2$ and no molecular gas detection) as a reference value and discuss the relative \avqpah\ in each gas phase. We find that the PAH fraction in the LMC diffuse neutral medium (4.1~\%) is similar to that of the MW diffuse neutral ISM, while in the SMC it is substantially lower than in the MW (1.1~\%). 
The galaxy-averaged \avqpah\ are consistent with both a power-law dependence of \qpah\ with metallicity \citep[][]{RemyRuyer15}, as well as the existence of a threshold around $12+{\rm log(O/H)} \sim 8.1$ at which the PAH abundance changes rapidly \citep[e.g.,][]{Draine07}. But \avqpah\ in the diffuse neutral gas favors the latter hypothesis.

We find evidence that \avqpah\ is systematically low in all identified \hii\ regions (Figures \ref{FigMCsqPAHImg}, \ref{FigMCsGasCorr}). Additionally, we show that the sphere of influence of the \hii\ regions on the PAH fraction grows as one would expect for a Str\"omgren sphere (Figure \ref{FigRadiiInfluence}). 

We investigate possible metallicity-dependent PAH formation and destruction scenarios to explain the higher PAH fraction in the neutral medium of the LMC compared to the SMC:
\begin{itemize}
    \setlength\itemsep{0.1em}
    \item[-] We find higher \avqpah\ in molecular gas with respect to the galaxy global averages, but similar to those of the diffuse neutral medium. This is consistent with formation of PAHs in the molecular gas and injection in the diffuse neutral medium, or vice-versa.
    \item[-] In each galaxy, we find a clear trend of \qpah\ with \halpha\ emission on global scale (Figure \ref{FigMCsGasCorr}). We find a limit in \halpha\ luminosity (or ionized gas fraction) at which the PAH abundance starts to decrease, and interpret this as a destruction of PAHs in the ionized gas. This limit is lower than the typical \halpha\ emission in an \hii\ regions, suggesting that the ionized medium, even outside of \hii\ regions, still affects the PAH fraction through destruction processes.
    \item[-] A radiation field intensity about twice that of the solar neighborhood seems to be enough to affect the PAH fraction, in both the neutral and ionized gas phase (Figure \ref{FigUbarqPAH}). However, the ionized medium always shows a lower PAH fraction than the neutral medium, even at equal intensity of the radiation field. 
    \item[-] We find that formation of PAHs through the fragmentation of large grains is a plausible explanation for the higher \avqpah\ in the LMC compared to the SMC diffuse neutral medium.
\end{itemize}

Future work will investigate the variations of the PAH fraction with metallicity in resolved, nearby galaxies. The launch of {\it the James Webb Space Telescope} will allow for the detection of PAHs at higher redshifts, and give us the opportunity to study the PAH fraction at low metallicity in the early Universe.

\acknowledgments{We are very grateful to the referee for their careful reading and their comments which greatly improved the paper. JC wishes to thank Jean-Philippe Bernard, Olivier Bern\'e and D\'eborah Paradis for fruitful discussions. The work of JC, KS, IC, AKL, and DU is supported by NASA ADAP grants NNX16AF48G and NNX17AF39G and National Science Foundation grant No.~1615728. The work of AKL and DU is partially supported by the National Science Foundation under Grants No.~1615105, 1615109, and 1653300. This paper used the Southern H-Alpha Sky Survey Atlas (SHASSA), which is supported by the National Science Foundation.}


%% copy editing.  Individual instruments can be provided in parentheses,
%% after the keyword, but they will not be verified.
\facility{Herschel, Spitzer}



\bibliographystyle{aasjournal}
\bibliography{pahs_mcs_refs}

\clearpage

\end{document}